\documentclass[twocolumn]{aastex631}

\newcommand{\numSimChamCont}{25}
\newcommand{\numSimChamBimod}{43}
\newcommand{\numSimUeda}{43}

\usepackage{todonotes} 
\usepackage{amsmath} 
\usepackage{multirow}
\usepackage{booktabs}
\let\tablewidth\relax
\usepackage{tabularray}
\usepackage{tabularx} 
\usepackage{ragged2e}   
\usepackage{natbib}     
\usepackage{wasysym}
\usepackage{xspace}
\usepackage{CJK}

\begin{document}
\begin{CJK*}{UTF8}{gbsn}

\title{Origins of Mercury's Big Heart of Iron: Exploring Pathways to Form High Core Mass Fraction (CMF) Planets via N-body Simulations}

\author[0000-0002-0786-7307]{Haniyeh Tajer}
\affiliation{Department of Astronomy, The Ohio State University, 100 W 18th Ave, Columbus, OH 43210 USA}

\author[0000-0002-4361-8885]{Ji Wang (王吉)}
\affiliation{Department of Astronomy, The Ohio State University, 100 W 18th Ave, Columbus, OH 43210 USA}

\author[0000-0002-9343-8612]{Anna C. Childs}
\affiliation{Department of Physics \& Astronomy, University of Alabama, Tuscaloosa, AL, 35487, USA}

\author[0000-0001-8140-1727]{Noah Ferich}
\affiliation{Department of Physics \& Astronomy, University of Nevada, Las Vegas, 4505 S. Maryland Pkwy, Las Vegas, NV 89154 USA}

\author[0000-0003-0834-8645]{Tiger Lu (陆均)}
\altaffiliation{Flatiron Research Fellow}
\affiliation{Center for Computational Astrophysics, Flatiron Institute, 162 5th Avenue, New York, NY 10010, USA}
\email{tlu@flatironinstitute.org}

\author[0000-0003-1927-731X]{Hanno Rein} 
\affiliation{Department of Physical and Environmental Sciences, University of Toronto at Scarborough, Toronto, Ontario M1C 1A4, Canada}
\affiliation{David A. Dunlap Department of Astronomy and Astrophysics, University of Toronto, Toronto, Ontario, M5S 3H4, Canada}
\email{hanno.rein@utoronto.ca}




\begin{abstract}

Mercury's core mass fraction (CMF) is ~0.7, more than double that of the other rocky planets in the solar system, which have CMFs of ~0.3. The origin of Mercury's large, iron-rich core remains unknown. Adding to this mystery, an elusive population of “Exo-Mercuries” with high densities is emerging. Therefore, understanding the formation of Mercury and its exoplanetary analogs is essential to developing a comprehensive planet formation theory. Two hypotheses have been proposed to explain the high CMF of Mercury: (1) giant impacts during the latest stages of planet formation strip away mantle layers, leaving Mercury with a large core; and (2) earlier-stage iron enrichment of planetesimals closer to the Sun leads to the formation of an iron-rich planet. In this work, we conduct N-body simulations to test these two possibilities. Our simulations are focused on the solar system, however, we aim to provide a framework that can later be applied to the formation of high-CMF exoplanets. To investigate the giant impact scenario, we employ uniform initial CMF distributions. To address the other hypothesis, we use a step function with higher CMFs in the inner region. For a uniform initial CMF distribution, our results indicate that although erosive impacts produce iron-rich particles, without mechanisms that deplete stripped mantle material, these particles merge with lower-CMF objects and do not lead to Mercury's elevated CMF. However, a step function initial CMF distribution leads to the formation of a high-CMF planet alongside Earth-like planets, resembling the architecture of the terrestrial solar system.

\end{abstract}


\section{Introduction} \label{sec:intro}

Mercury's high core mass fraction (CMF), relative to the other terrestrial planets in the solar system, remains a mystery to this day. Earth, Mars, and Venus each have a CMF of around 0.3, whereas Mercury's CMF is more than double, at approximately 0.7 \citep{SOHL201523, adibekyan_compositional_2021, Taylor_McLennan_2008}. The evolutionary pathway that led to this discrepancy is not well understood. One dominant theory suggests that a giant impact stripped away the outer mantle layers of proto-Mercury, leaving it with an oversized core \citep{benz_collisional_1988, benz_origin_2007}. However, observations of volatile elements on the surface challenge this theory \citep{Peplowski2011}. Moreover, simulations have shown that a single giant impact is unlikely to be energetic and efficient enough to remove sufficient mantle material to result in such a high core mass fraction \citep{10.1093/mnras/stae644, franco_explaining_2022}. In contrast to giant impacts, gradual mantle stripping due to many minor erosive impacts is an alternative way of increasing the CMF \citep{hyodo_modification_2020, chau_forming_2018}. Because the gravitational field is weaker and facilitates mantle loss, \citet{hyodo_modification_2020} argue that collisions during the planetesimal stage create iron-rich planetesimals by eroding their mantles, and these planetesimals later merge to form Mercury.

In the absence of giant and minor impacts, some studies propose alternative physical processes that may lead to the formation of iron-rich planetesimals in the inner regions of the protoplanetary disk. These metal-rich planetesimals can subsequently form Mercury-like planets with high core mass fractions. One such theory involves iron-rich nucleation and its separation from silicates before pebble formation. \citet{johansen_nucleation_2022} argue that, due to the high surface tension of iron relative to silicates, iron pebbles stick together more easily and grow larger in the inner regions of the protoplanetary disk, where temperatures are higher. Consequently, the presence of larger iron pebbles would lead to the formation of iron-rich planetesimals, which could later contribute to the formation of iron-rich planets.

\citet{kruss_seeding_2018} propose another plausible scenario in which magnetic fields facilitate the aggregation of iron particles, from dust grains to particles larger than millimeters in size. During the early stages of planetesimal formation, dust grains attempt to coalesce into larger aggregates, but they are hindered by bouncing rather than fragmentation, facing the ``bouncing barrier" \citep{zsom_outcome_2010}. Since silicates are not affected by magnetic fields, they struggle to grow larger than millimeter-sized particles and fail to overcome the bouncing barrier, whereas iron-rich aggregates can form larger particles under the influence of a magnetic field \cite{kruss_seeding_2018, kruss_composition_2020}.

Photophoresis can also play a role in Mercury's formation by separating iron-rich material and silicates due to different thermal conductivities. \citep{Wurm2018, wurm_photophoretic_2013}.

All of these processes contribute to the formation of iron-rich planetesimals in the inner regions of the proto-solar system. 


\begin{figure}
    \centering
    \includegraphics[width=1.0\linewidth]{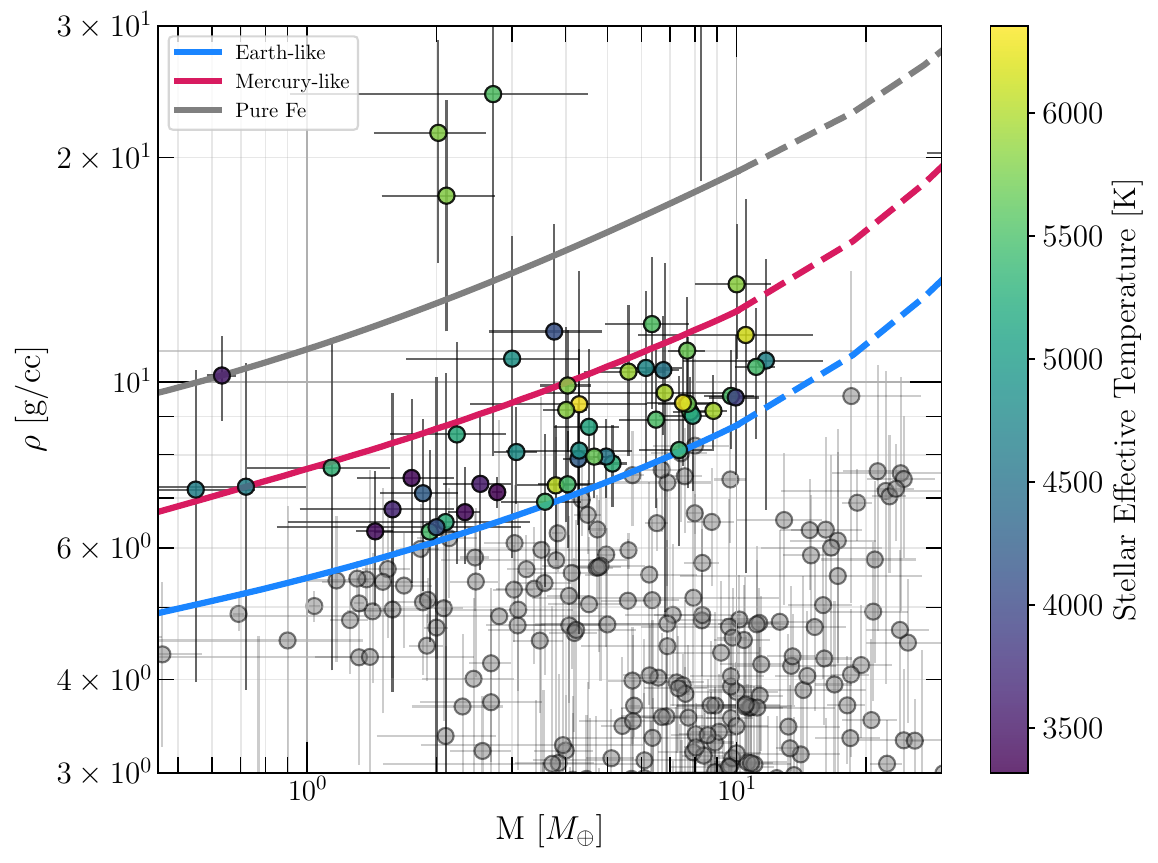}
    
    \caption{Mass-density relationships for a sample of exoplanets, including exo-Mercuries.
    We derive this sample from the NASA exoplanet archive \citep{christiansen_nasa_2025}, only keeping planets that have mass, radius, and stellar effective temperature measurements.
    The blue, red, and black curves are mass-density relationships for a planet with Earth-like, Mercury-like, and pure iron composition. These curves are derived from planetary interior models made with \texttt{ExoPlex} \citep{Unterborn_2023}.
    We colored planets whose measured bulk densities were higher than the predicted Earth-like density given their mass. Colors represent the stellar effective temperature. The gray points are all the other planets.
    }
    
    \label{fig:exo-mercs}
\end{figure}

Interestingly, Mercury is not alone in this mystery. With advancements in our rocky exoplanet detection capabilities, a population of extremely dense exoplanets is on the rise. These exoplanets are dubbed ``Exo-Mercuries" or ``Super-Mercuries" \citep{valencia_internal_2006}. Their bulk densities imply very high core-mass fractions, more consistent with a Mercury-like composition than an Earth-like one \citep[e.g., CMF of 0.7 for Kepler-107 c][]{schulze_probability_2021}. Fig. \ref{fig:exo-mercs} shows some exo-Mercury candidates \citep{johansen_nucleation_2022, adibekyan2022, schulze_probability_2021, adibekyan_compositional_2021,toledo-padron_characterization_2020, bonomo_giant_2019, santerne_earth-sized_2018}. 

Looking at exoplanetary analogs of Mercury, they are diverse and reside in systems with various orbital architectures. For example, not all exo-Mercuries are the closest planet to their host star, e.g., Kepler-107~c \citep{bonomo_giant_2019}. This high-density planet has an orbital period of nearly 5 days \citep{bonomo_giant_2019}; however, there is an inner planet companion, Kepler-107 b, with an orbital period of almost 3 days. Kepler-107 b has a mass and radius consistent with an Earth-like model, whereas Kepler-107 c falls in the Mercury-like region \citep{schulze_probability_2021}.  Not all of these systems share the odd system structure of Kepler-107, e.g., K2-38 b is the innermost planet to the star, with an extremely high density of $\approx 9 \ g/cm^3$.

As a first step to answer all these questions about the formation of Mercury and its analogs, we start with separating the two major camps of formation theories: 1) Formation of Mercury from an iron-rich region, made possible by some astrophysical phenomena that have made the inner disk more iron-enriched, 2) Stochastic impacts between planetesimals of the same composition, leading to erosive collisions that strip away mantle material \citep[e.g. giant impact, or a hit-and-run collision][]{franco_explaining_2022, chau_forming_2018, benz_collisional_1988}

To this end, we perform N-body simulations of the late stages of planet formation, after planetesimals and planetary embryos have formed. At this stage, bodies are differentiated, so we can test different initial planetesimal compositions to compare the two camps. In this paper, we use the state-of-the-art N-body code \texttt{REBOUND} \citep{rebound2012} to simulate the late stages of planet formation, where collisional fragmentation takes place.

Previous studies have shown that assuming all collisions between planetesimals result in perfect mergers neglects important implications of fragmentation, where erosive impacts generate smaller fragments \citep{Chambers2013}. Accordingly, there have been efforts to include collisional fragmentation in simulations, such as those done by \citet{Chambers2013}. In their work, they also implement a mantle stripping model; however, their study is limited to a uniform initial CMF distribution of 0.3 for all bodies. Therefore, it does not address situations where the disk is differentiated and has a changing composition with respect to distance from the star. \citet{Clement2019} conduct a similar study, where they run N-body simulations for three different disks. The first set of their simulations is similar to \citet{Chambers2013} with minor differences. Their second set is similar to the first, but they introduce extra instability to the system by adding more giant planets (Jupiter and Saturn are already included in the first set). Their final set is a narrow annulus, based on the Grand Tack scenario \citep{Walsh2011}. They also implement a mantle stripping model, but they too only consider an initial uniform CMF distribution of 0.3 for all bodies. Moreover, in their simulations, they ignore gravitational interactions between smaller bodies, which can neglect important fragmentation effects. Even though they are successful in building Mercury in their simulations, their results demonstrate a low probability of Mercury's formation. Additionally, their work is focused on reproducing the solar system, and specifically the solar system's Mercury. In contrast, our work does not claim to reproduce the solar system with its precise characteristics of the four terrestrial planets and the relationships between their orbital parameters. Even though we are using initial conditions from solar system simulations, our aim is to develop a framework that we can later apply to exo-Mercury formation studies.


To do so, we build upon previous efforts in studying the probability of forming a Mercury-like planet via N-body simulations. We use the fragmentation module developed by \cite{childs_collisional_2022} to account for different possible collision outcomes, from mergers to catastrophic disruption. Our work differs in a few key points: 1) we experiment with the initial CMF distribution to study the effects of changing disk composition; 2) we experiment with different initial surface density distributions, specifically with the model derived by \citet{ueda_early_2021}, 3) our fragmentation module accounts for collisions between small objects---we have a minimum fragment mass that allows for fragmentation to take place between small planetesimals.

The structure of the paper is as follows: In \S\ref{sec:methods}, we describe the logic of our simulations, collision prescription, and mantle stripping model. In \S\ref{sec:results}, we present our results of testing different initial CMF and surface density distributions. In \S\ref{sec:discussion}, we discuss future work and limitations. In \S\ref{sec:conclusions}, we present our conclusions.

\section{Methods} \label{sec:methods}

\subsection{N-body simulations and collision prescription} \label{subsec: N-body simulations}

To conduct our simulations, we use the open-source N-body code \texttt{REBOUND} \citep{rebound2012}\footnote{\url{https://rebound.readthedocs.io/en/latest/}}. Since \texttt{REBOUND} does not have a built-in fragmentation module, we adopt the one developed by \citet{childs_collisional_2022}\footnote{\url{https://github.com/annacrnn/rebound_fragmentation}}, which is based on the results from hydrodynamic simulations presented by \citet{Leinhardt2012} \citep[also see][]{Chambers2013}. We refer readers to the mentioned papers for more details on the collision prescription, but here we briefly mention some key points.

Within this model, the collision outcome depends on the masses of the bodies (target mass $M_t$, projectile mass $M_p$), the impact velocity $v_i$, and the impact angle $\theta$. For high-energy head-on collisions where the impact velocity is higher than the mutual escape velocity, an erosive collision might happen. However, we need to note the computational limitation imposed by the minimum fragment mass allowed (see Table~\ref{tab:simulations_info} for values). In our simulations, an erosive collision can only happen if the resulting fragments fall above the minimum fragment mass threshold. Otherwise, depending on the impact angle, the two objects will either leave each other intact (``elastic bounce") or will merge. We experimented with a smaller minimum fragment mass (half of the values listed in \ref{tab:simulations_info}), and our results did not change. However, changing the minimum fragment mass by a few orders of magnitude might have more severe effects. We discuss this in \S\ref{sec:min_frag_mass}.

In general, if the impact velocity is less than the escape velocity, the particles will merge. Another possible scenario is where the impact angle $\theta$ is such that the impact parameter $B$, defined as $B = \sin\theta (R_t + R_p)$, is higher than the critical value $R_t$, where $R_t$ and $R_p$ are the radii of the target and projectile, respectively. This case can either cause a ``grazing partial erosion" of the target or a ``hit-and-run" event where some of the projectile's mass is eroded and absorbed by the target. Details of collision outcomes can be found in \citet{childs_collisional_2022}.

For all simulations, we use the hybrid integrator MERCURIUS \citep{rein_hybrid_2019}. Our timestep is 6 days, which is less than 10\% of the smallest orbital period of the planetesimals in all simulation sets (i.e., 66 days). Our simulations have a relative mass error of $(M - M_{0})/M_0 \approx 10^{-8}$, where $M$ represents the total mass in the simulation at different time points, and the subscript 0 denotes the initial mass. 



\subsection{Initial conditions for planetesimals} \label{subsec: Sigma}

\begin{splitdeluxetable*}{lccccBccccc}
\tablecaption{Summary of initial conditions used in each simulation set \label{tab:simulations_info}}
\tablehead{
\colhead{Simulation name} & \colhead{Number of converged runs} & \colhead{Total disk mass ($M_{\oplus}$)} & \colhead{Number of bodies} & \colhead{Mass of particles ($M_{\oplus}$)} & \colhead{Disk's inner edge (au)} & \colhead{Disk's outer edge (au)} & \colhead{Minimum fragment mass ($M_{\oplus}$)} & \colhead{Reference}
}
\startdata
\textit{Ueda 2021} & \numSimUeda & 2.200 & 150 & 0.015 & 0.40 & 1.38 & 0.0075 & \citet{ueda_early_2021}\\ 
\textit{Chambers 2013 bimodal} & \numSimChamBimod & 2.604 & 154 & 0.01 - 0.1 & 0.30 & 2.00 & 0.005 & \citet{Chambers2013} \\ 
\textit{Chambers 2013 continuous} & \numSimChamCont & 2.536 & 154 & Varying with orbital distance & 0.30 & 2.00 & 0.005 & \citet{Chambers2013} \\ 
\enddata
\tablecomments{See \S~\ref{subsec: Sigma} for a detailed description of initial conditions used for each simulation, including the surface density distributions of planetesimals.}
\end{splitdeluxetable*}

We conduct three sets of simulations, two with the initial conditions from \cite{Chambers2013}, and one based on the model from \cite{ueda_early_2021}. All simulations are integrated for 300 Myr and include the Sun. 

In the first set (\textit{Chambers 2013 bimodal}), we start with a disk of bimodal mass distribution: 14 Mars-sized planetary embryos ($0.1 M_{\oplus}$) and 140 moon-sized ($0.01 M_{\oplus}$) planetesimals. The inner and outer edges of the disk are 0.3 and 2.0 au. The surface density distribution is different between the inner and outer disk. In the inner disk, surface density increases linearly from 0.3 to 0.7 au. In the outer disk, from 0.7 to 2 au, surface density distribution follows a power law $a^{-3/2}$, normalized to 8 $\text{g/cm}^2$ at 1 au. The radii of all objects are calculated assuming a density of 3 $\text{g/cm}^3$.

We choose the initial inclination $i$, eccentricity $e$, longitude of ascending node $\Omega$, argument of pericenter $\omega$, and true anomaly $f$ in the same manner as \cite{childs_collisional_2022}, from a random uniform distribution, where $0 < e < 0.01$, $0 < i < 0.0175$, $0 < \omega < 2 \pi$, $0 < \Omega < 2\pi$, and $0 < f < 2\pi$. 

In all simulations, we add Jupiter and Saturn in their current orbits to perturb the orbits of other particles, to be more representative of the solar system architecture, as well as to induce more frequent collisions.

The second simulation set is similar to the first set, except for the planetesimal and embryo masses. This set is also described in \citet{Chambers2013}. Here, we have 14 planetary embryos and 140 planetesimals, however, the mass of each object follows the surface density distribution described in \textit{Chambers 2013 bimodal}.

Our third set is based on the model presented in \cite{ueda_early_2021}. Instead of a surface density distribution that follows a power-law, \citet{ueda_early_2021} provide evidence for a ring-like planetesimal distribution. They conduct simulations of gas and dust evolution and find a planetesimal surface density profile with a high concentration of mass in near Earth's orbit, and a mass deficit in Mercury and Mars's regions. This potentially explains the low-mass Mars problem. In this work, we test one of the proposed surface density profiles in \cite{ueda_early_2021}, with a total mass of $2.2 M_{\oplus}$, which corresponds to a turbulence of $\alpha_{dead} = 2.3 \times 10^{-4}$. Specifically, we replicate their proposed surface density distribution, which is similar to a narrow ring peaking at 0.78 au, which is equivalent to $r_{peak}$ in \cite{ueda_early_2021}. The disk's inner and outer edges are at 0.4 and 1.38 au, respectively.

To be comparable with the other two simulation sets, we set the number of bodies to 150 bodies, each with a mass of $0.015 M_{\oplus}$. We also set the minimum allowed fragment mass as half the mass of planetesimals. Other orbital parameters, such as eccentricity and inclination, are randomly chosen in the same manner as for the other two simulation sets. Similarly, we include Jupiter and Saturn in these simulations.

\subsection{Mantle stripping model} \label{subsec: mantle stripping}
\cite{ferich2024} developed a post-processing tool, the ``Differentiated Body Composition Tracker" (DBCT) \footnote{\url{https://github.com/Nofe4108/Differentiated_Body_Composition_Tracker}}, that works in conjunction with the fragmentation module from \cite{childs_collisional_2022}. DBCT takes the collision history and initial CMF values as input and iterates through collisions to track the change in composition of bodies involved in these events.

We build upon their work to develop a post-processing tool that takes the same input but uses a different mantle stripping model. We use a model described in \cite{marcus_watericy_2010}. They present two mantle stripping models and compare their differences, ultimately showing that there is not a significant difference between the outcomes of the two models.

The first model takes the escaping mass from the mantle material first. To start, we compute the total core mass available:

\begin{equation}
\label{core_tot}
M_{core, tot} = CMF_{t} \  M_{t} + CMF_{p} \   M_{p}
\end{equation}
where $M_{core, tot}$ is the total core mass, and subscripts $t$ and $p$ refer to target and projectile respectively. 

If, in an erosive collision, mass of the largest remnant $M_{lr}$ is less than or equal to the total core mass available, then the CMF of the largest remnant will be 1, and the rest of the core material is distributed evenly between fragments.

\begin{equation}
\label{eq:cmf_lr_is_one}
CMF_{lr} = 1
\end{equation}
\begin{equation}
\label{eq:cmf_frag_both}
CMF_{frag} = \frac{M_{core, tot} - M_{lr}}{M_r}
\end{equation}
Where $CMF_{frag}$ is each fragment's CMF, $CMF_{lr}$ is CMF of the largest remnant, and $M_r$ is the remaining mass:
\begin{equation}
M_r = M_t + M_p - M_{lr}
\end{equation}

On the other hand, if $M_{lr}$ is larger than the total core mass available, then:

\begin{equation}
\label{cmf_lr}
CMF_{lr} = \frac{M_{core, tot}}{M_{lr}}
\end{equation}
\begin{equation}
CMF_{frag} = 0
\end{equation}

So all fragments are mantle material only in this case.

In the second model, in an erosive event, mantle material is ejected from the target until there is no more mantle left, and then the target's core is ejected. In other words, if the mass of the largest remnant is bigger than the target's core mass, then:

\begin{equation}
\label{cmf_lr_model2_case1}
CMF_{lr} = \frac{M_{core, target}}{M_{lr}}
\end{equation}

\begin{equation}
CMF_{frag} = \frac{M_{core,tot} - M_{core, target}}{M_r}
\end{equation}

If not (i.e., $M_{lr} < M_{core, target}$), then particles are treated the same as model 1, following equations \ref{eq:cmf_lr_is_one} and \ref{eq:cmf_frag_both}.


We analyze our results using both models and find that the differences between their outcomes are negligible, consistent with the findings of \citet{Marcus2010}. By the time planets fully evolve, fragments either merge with one another or accrete onto larger bodies, so minor variations in their computed CMFs are effectively washed out. These differences may become more significant under alternative conditions (e.g., a denser disk or additional mechanisms that promote more efficient radial mixing); however, they remain negligible for standard solar system studies. Therefore, we present our results using only the first model.

There is a caveat for following CMF changes with a post-processing module. Changes in CMFs result in variations in densities, hence changes in radii. This is inconsistent with the assumption of constant density for all particles throughout the N-body simulations. However, modifications in radii in this small domain do not significantly affect the simulation results. \citet{childs_collisional_2022} study the effects of using expansion factors, where they inflate the bodies to reduce collision timescales while conserving the body masses. Their results show no significant differences in simulation outcomes for an expansion factor of up to 3, which is much larger than the changes in radii due to CMF changes.

\subsection{Initial CMF distributions} \label{subsec:CMF_i}
In this work, we examine two opposing scenarios: 
\begin{enumerate}
    \item Erosive collisions between massive planetesimals can create an iron-rich planet with a CMF as high as Mercury's.
    \item Mercury was born from an \textit{a priori} iron-rich inner region of planetesimals, distinct from other regions in the solar system.
\end{enumerate}

We can only test the first scenario for collisions between massive planetesimals. \citet{hyodo_modification_2020} show that mantle loss is more efficient for lower mass objects, but erosive collisions can still happen closer to the star for massive planetesimals. Therefore, the results of \citet{hyodo_modification_2020} will fit in our second scenario, as they might make an iron-rich inner region of planetesimals. We discuss the connection between our simulations and \citet{hyodo_modification_2020} in \S~\ref{sec:hyodo}. 

Other processes may lead to iron-enrichment of the inner region, such as magnetic aggregation \citep{kruss_seeding_2018}, effects of surface tension \citep{johansen_nucleation_2022}, and photophoresis \citep{Wurm2018}. These examples are discussed in \S~\ref{sec:intro}.

To test the two scenarios (stochastic giant impacts versus an iron-rich inner region), we vary the initial CMF distributions in the disk. If the first scenario is correct, then an iron-rich planet can form as a result of collisions between massive bodies of the same composition. Therefore, we use uniform initial CMF distributions for this case. We should also note that we use high-CMF and iron-rich interchangeably, as we assume a planet's core is entirely composed of iron.

To test the second scenario (i.e., Mercury's formation from a distinct iron-rich region), we use a step function in the initial CMF distribution, assigning higher iron content to the inner parts of the disk. To conserve the total iron content of the disk, we trade the excess iron in the inner disk for a deficit of iron in the outer disk. Consequently, the outer disk will have a CMF lower than the average CMF of the entire disk. Accordingly, we examine whether planets with Earth-like CMFs can still form, rather than being extremely iron-depleted. Our CMF choices for the step functions are listed in Appendix \ref{sec:app_tables} and explained further in \S~\ref{subsec:step_func}.

\subsubsection{Uniform initial CMFs and the effects of metallicity}\label{subsubsec:uniform_initial_cmfs_and_metallicity}

\begin{figure}
    \centering
    \includegraphics[width=1\linewidth]{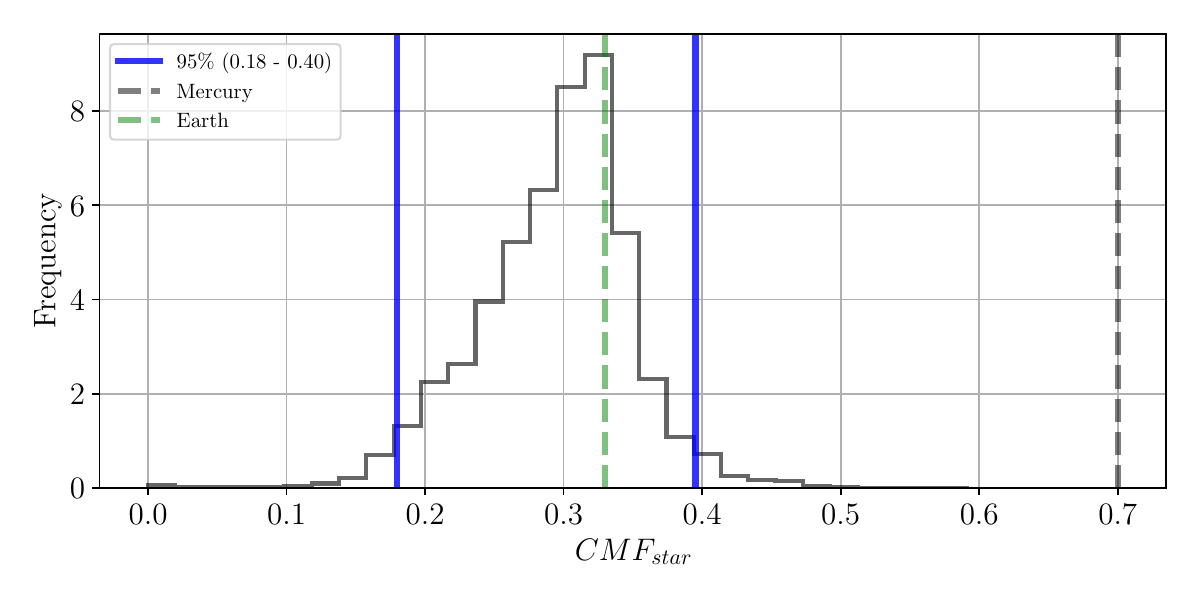}
    \caption{Histogram of inferred $CMF_{star}$ values for nearby stars, based on abundances from Hypatia Catalog \citep{hinkel_stellar_2014}. Blue solid lines indicate where 95\% of the data falls. We choose the upper bound of the 95\% interval ($CMF_{star} = 0.40$) for our metallicity test. Green and grey dashed lines depict CMF values for the Earth and Mercury, respectively. }
    \label{fig:hypatia_hist}
\end{figure}

For uniform initial CMF distributions, we first choose a CMF value of 0.33 for all particles to reflect an Earth-like composition.

We also explore the effects of stellar metallicity. In general, rocky planets reflect the elemental abundances of their host star \citep{schulze_probability_2021}. Therefore, for a higher metallicity star, we expect to have denser, more metal-rich planets. We derive $CMF_{star}$ as described by \cite{schulze_probability_2021}:
\begin{equation}
\label{cmf_star}
CMF_{star} = \frac{(\frac{Fe}{Mg})m_{Fe}}{(\frac{Fe}{Mg})m_{Fe} + (\frac{Si}{Mg})m_{SiO_2} + m_{MgO}}
\end{equation}
We compute $CMF_{star}$ for nearby stars with measured abundances from the Hypatia catalog \cite{hinkel_stellar_2014}. Figure \ref{fig:hypatia_hist} shows a histogram of these values. 95\% of all derived $CMF_{star}$ values fall between 0.18 and 0.40, with a mean value of 0.34. Therefore, we also test a uniform initial CMF distribution of 0.40 in addition to 0.33 to study the effects of higher metallicity on the compositions of simulated planets.

\subsubsection{Step function initial CMFs}\label{subsec:step_func}

Multiple physical processes can make the inner regions of the disk more iron-rich, as described in \S~\ref{sec:intro}. To test this, we conserve the total mass of core material in the disk while changing the amounts of iron in the inner and outer regions of the disk. We vary the boundary ($b$) and the percentage of exchanged iron material ($x$) between the inner and outer region. Details of these choices are noted in tables in Appendix \ref{sec:app_tables}. In short, We test for boundaries between 0.5 and 0.8 au, with different percentages of iron exchange, varying between 10\% to 40\%. Assuming a boundary $b$, we first calculate the disk mass that resides in the inner region ($r < b$) and the outer region ($r \ge b$). Next, we calculate the total mass of core material in the outer region:
\begin{equation}
M_{core, outer} = CMF_{total} \ M_{outer}
\label{m_core_outer}
\end{equation}
where $CMF_{total}$ is the total disk CMF based on $CMF_{star}$. Then, based on the chosen percentage of iron exchange ($x$), we compute the amount of iron to be brought to the inner disk:
\begin{equation}
M_{exchange} = x \ M_{core, outer}
\end{equation}
We then replace this mass in the inner disk with an equal mass of mantle material, thus keeping the total mass of the inner disk constant. The replaced mantle material is moved to the outer disk, also conserving the outer disk's total mass. We then compute the resulting CMFs for the inner and outer disk:
\begin{equation}
 CMF_{inner} = \frac{CMF_{total}M_{inner} + xM_{core, outer}}{M_{inner}}
\label{eq:cmf_inner}
\end{equation}
\begin{equation}
\begin{split}
CMF_{outer} &= \frac{CMF_{total}M_{outer} - xM_{core, outer}}{M_{outer}} \\
&= (1 - x) CMF_{total}
\end{split}
\label{eq:cmf_outer}
\end{equation}

According to Equation \ref{eq:cmf_outer}, the value of $CMF_{outer}$ is directly related to the percentage of iron that is being taken away. Therefore, to have systems that produce both Mercury and Earth-like planets, $x$ can not approach 100\%. Therefore, we add an additional variable, boundary $b$, to control the iron exchange between the inner and outer disk. 
 
\section{Results} \label{sec:results}

In this section, we present our results. \S~\ref{sec:uniform_results} describes simulations with a uniform initial composition. \S~\ref{sec:step_results} covers simulations with a step-function initial disk composition. Finally, in \S~\ref{sec:combine_results}, we compare the uniform and step-function results and draw our conclusions.

\subsection{Uniform initial composition}
\label{sec:uniform_results}

\begin{figure*}
    \centering
    \includegraphics[width=1\linewidth]{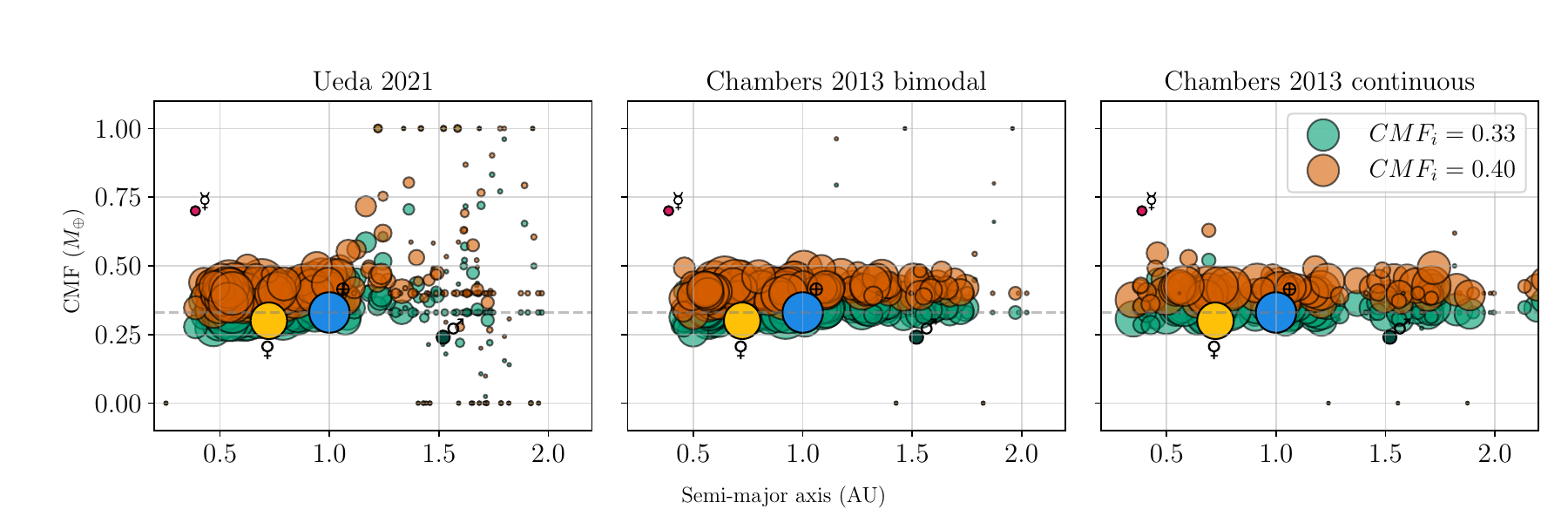}
    \caption{
    Final planets' CMFs versus semi-major axis for two different uniform initial compositions: orange points have an initial CMF ($CMF_i$) of 0.33, and teal ones start with $CMF_i = 0.40$. Point sizes are scaled to match their masses, and solar system planets are plotted for reference. Each plot depicts results for a different set of initial conditions, detailed in Table~\ref{tab:simulations_info}.
    Planets evolved from higher initial CMFs move slightly upward in CMF values, but not high enough to reach the CMF of Mercury. The \textit{Ueda 2021} simulation set shows many small high- and low-CMF objects scattered in the outer disk regions, which are discussed in \S\ref{sec:diss_init_conds}.}
    
    \label{fig:high-metal-all-three}
\end{figure*}

We test a uniform initial composition for the disk, assuming that all planetesimals start with the same CMF. We adopt a 0.33 CMF value representing an Earth-like composition for the planetesimals. We also test a higher initial CMF (0.40) to see whether this choice will help with enhancing the CMFs of the simulated planets (See \S~\ref{subsubsec:uniform_initial_cmfs_and_metallicity}). As shown in Fig.~\ref{fig:hypatia_hist}, 0.40 is the upper limit for 95\% of CMFs inferred from nearby star metallicities. Accordingly, we use an initial CMF value of 0.40 for all bodies and examine the resulting planets.

We plot the CMFs of simulated planets against their final semi-major axes and compare them with the solar system planets in Figure \ref{fig:high-metal-all-three}. Orange and teal circles represent the final planets from the simulations, with sizes scaled to their masses. The solar system planets are marked with their respective Roman letters and are also scaled to represent their respective masses. Each plot refers to a certain set of initial conditions, noted in Table~\ref{tab:simulations_info}. We discuss the effects of initial planetesimal mass distributions in \S~\ref{sec:diss_init_conds}. For each set of initial conditions, results from all converged simulations are plotted together.

Mercury stands out in the high-CMF region close to the star. No simulated planet falls in that region, regardless of the choice of initial planetesimal mass distribution. In simulations with \textit{Ueda 2021} initial conditions, we observe many high- and low-CMF objects with relatively low mass, scattered in the outer disk regions, but not in Mercury's proximity. We discuss possible explanations for this in \S~\ref{sec:diss_init_conds}. 

We conclude that with a higher initial CMF, simulated planets' CMFs are slightly enhanced, but still cannot reach Mercury's level \citep{schulze_probability_2021, Unterborn_2023}. One reason for this is that giant impacts energetic enough to remove large amounts of mantle are extremely rare \citep{franco_explaining_2022}. Moreover, although erosive events occur in our simulations and produce high-CMF fragments (see \S~\ref{sec:combine_results}), our simulations lack a mechanism for removing low-CMF mantle materials. These low-CMF objects will merge with high-CMF fragments later to eventually renormalize all final planets to a moderate CMF. Forming Mercury from a uniform CMF distribution may therefore require a process that removes mantle material from the disk \citep{spalding_solar_2020}.

It is also important to note that mantle-removing collisions are more efficient for lower-mass bodies (see \S~\ref{sec:hyodo}). Other iron-enriching mechanisms can also occur at smaller particle stages~\citep{kruss_seeding_2018, Kruss2018, johansen_nucleation_2022, wurm_photophoretic_2013, Wurm2018}. Consequently, these mechanisms in the earlier stages of dust and planetesimal growth may lead to an inner planetesimal disk that is already enriched in high-CMF objects. We discuss this in the section below. 
 
\subsection{Step function CMF distributions}
\label{sec:step_results}

\begin{figure*}
    \centering
    \includegraphics[width=1\linewidth]{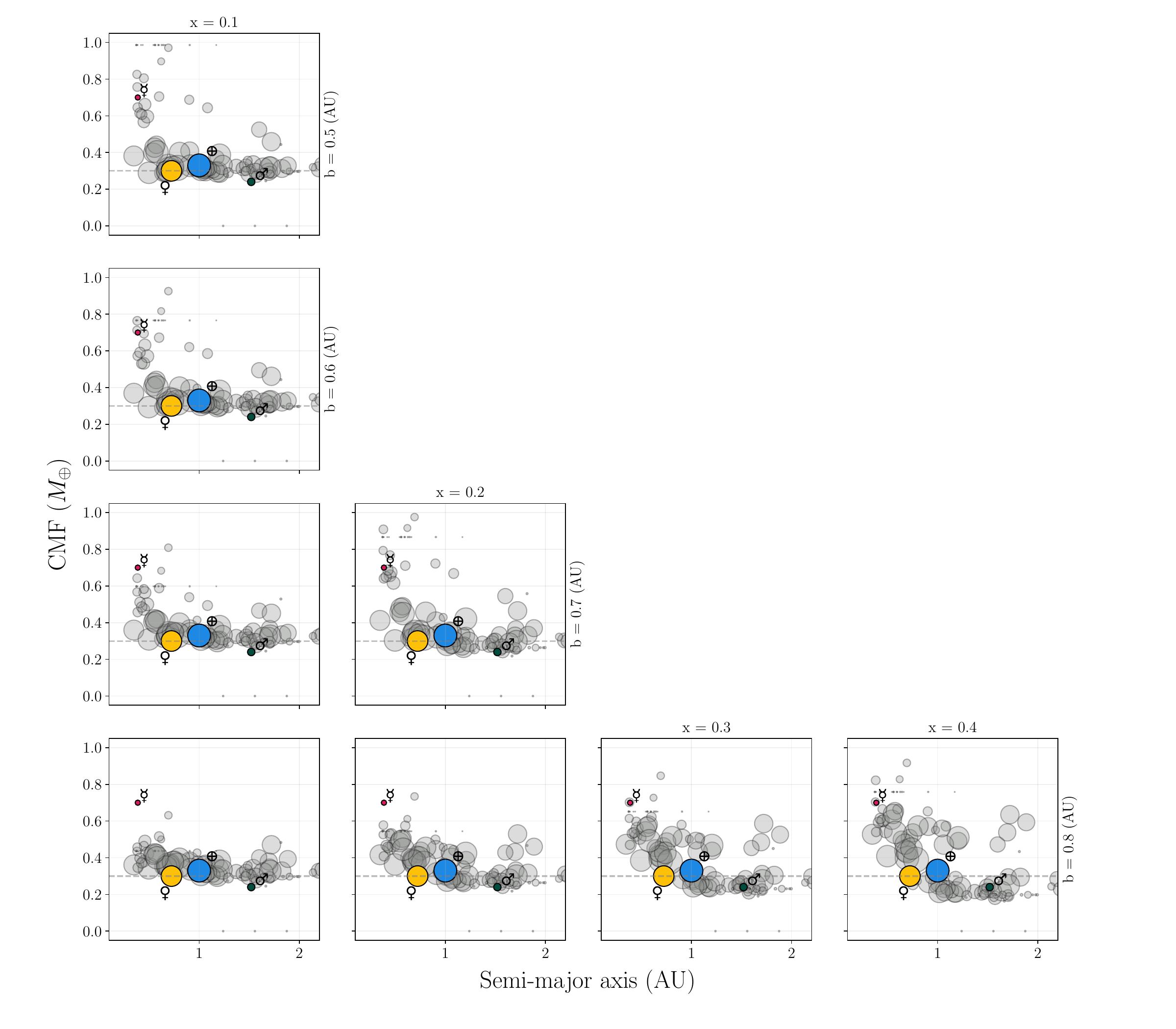}
    \caption{CMF versus semi-major axis for planets evolved from the \textit{Chambers 2013 continuous} simulation set. We explored different boundaries and amounts of iron exchange between the inner and outer disk regions. Each row corresponds to a different boundary, ranging from 0.5 to 0.8 au. Each column represents a different iron exchange percentage ($x$ in equation~\ref{eq:cmf_inner}), increasing from 10\% on the far left to 40\% on the far right. Details of this set of initial conditions are listed in Table~\ref{tab:simulations_info}.}
    \label{fig:grid_cham_cont}
\end{figure*}

\begin{deluxetable}{ccc}
\tablecaption{MDI values for different simulation sets \label{tab:mdi-all_three}}
\tablehead{
\colhead{Simulation name} & \colhead{Initial CMF distribution} & \colhead{MDI}}
\startdata
\textit{Ueda 2021} & Uniform & 0.301 \\
\textit{Chambers 2013 bimodal} & Uniform &  0.199 \\ 
\textit{Chambers 2013 continuous} & Uniform & 0.190 \\ 
\textit{Ueda 2021} & Step function &  0.285 \\ 
\textit{Chambers 2013 bimodal} & Step function &  0.082 \\
\textit{Chambers 2013 continuous} & Step function & 0.000 \\
\enddata
\tablecomments{ \textit{Chambers 2013 continuous} simulation set with a step function initial CMF distribution has the lowest MDI, therefore it is the most successful in reproducing Mercury's mass and CMF.}

\end{deluxetable}

Various studies have suggested mechanisms for enriching the inner planetesimal disk with iron (see \ref{sec:intro}).
Motivated by scenarios that alter the initial CMF distribution, we adopt a step-function to describe the CMF distribution for the initial disk, assigning higher CMFs to the objects in the inner region. We vary the inner-disk iron enrichment by adjusting the inner/outer disk boundary and the percentage of iron mass swapped. The inner/outer disk boundaries range from 0.5 to 0.8 au, and the percentage of iron mass swapped ranges from 10\% to 40\%. The corresponding results for the \textit{Chambers 2013 continuous} simulation set are shown in Fig.~\ref{fig:grid_cham_cont}. Details of the CMF choices are listed in appendix~\ref{sec:app_tables}. The results of the other two simulation sets are discussed in \S~\ref{sec:diss_init_conds}.

Fig.~\ref{fig:grid_cham_cont} depicts the impact of inner/outer disk boundary and iron swap percentage on the final outcome of the simulated planets. The sizes of the scatter points are scaled to match the masses of the simulated planets. Each row of plots corresponds to a different choice of boundary, and each column is for a different choice of $x$. Enhancing the initial CMFs of the objects in the inner disk leads to higher CMFs for the evolved planets closer to the star. There is also a scatter in the planets' CMFs further from the star (e.g., boundary of 0.8 au and $x = 0.4$). Most of these planets have formed from the inner disk region and then migrated outwards. This is consistent with the considerable amount of migration and scattering observed in our simulations. However, when planetesimals migrate inward, they are likely to merge with another particle, as inner regions have a higher particle number density. In contrast, if they migrate outwards, the particle number density is lower, hence the likelihood of interaction with another particle is lower.

To quantitatively compare simulation results along different choices of boundary and x, we define an Absolute Mercury Deviation Index, $MDI_{abs}$:

\begin{equation}
    MDI_{abs} = (\frac{\Sigma (CMF_{sim} - CMF_{\mercury})^2}{ave(CMF_{sim}).N} +
    \frac{\Sigma (M_{sim} - M_{\mercury})^2}{ave(M_{sim}).N})^{1/2}
    \label{eq:merc_dev_index}
\end{equation}

Where subscript $sim$ refers to simulated Mercuries and subscript $\mercury$ refers to the true values for Mercury. $N$ is the number of simulations. We define a simulated Mercury as any planet in the final system whose semi-major axis is smaller than 0.5 au.

We adjust MDI by subtracting the minimum value across all different setups so that the initial condition that produces planets that resemble Mercury the most has an MDI value of zero.

\begin{equation}
    MDI = MDI_{abs} - min(MDI)
\end{equation}

Here, for the step function initial CMF distributions, we compute MDI for all different choices of $b$ and $x$ together. Later, we will compute MDI for individual choices of $b$ and $x$.

\textit{Chambers 2013 continuous} simulation set with a step function initial CMF distribution has the lowest $MDI$ value (i.e., zero) and therefore is the most successful simulation set in producing planets with CMFs and masses closer to Mercury. Table~\ref{tab:mdi-all_three} lists the MDI values for different simulation sets. We find that step-function simulations better reproduce Mercury analogs with generally lower MDI values.

\begin{table}[htbp]
\label{tab:mdi-cham-cont}
\centering
\caption{MDI values for different choices of $b$ and $x$ for \textit{Chambers 2013 continuous} simulation set. \label{tab:mdi-cham-cont}}
\begin{tabular*}{\dimexpr 0.3\textwidth}{@{\extracolsep{\fill}} ccc @{}}
\hline\hline
b (au) & x & MDI \\
\hline
0.5 & 0.1 & 0.031 \\
0.6 & 0.1 & 0.013 \\
0.7 & 0.1 & 0.036 \\
0.7 & 0.2 & 0.013 \\
0.8 & 0.1 & 0.122 \\
0.8 & 0.2 & 0.051 \\
0.8 & 0.3 & 0.014 \\
0.8 & 0.4 & 0.000 \\
\hline
\end{tabular*}
\parbox{0.4\textwidth}{\footnotesize \vspace{2pt} 
MDI values are normalized by subtracting the minimum $MDI_{abs}$ value, which is for $b = 0.8 \ {au}$ and $x = 0.4$. Higher MDI values imply a weaker resemblance to Mercury's CMF.}
\end{table}

We calculate $MDI_{abs}$ for different choices of boundary and percentage of iron exchange in the \textit{Chambers 2013 continuous} simulation group. We choose this group because it has the lowest MDI value, compared to the other groups. Here, we normalize the values by subtracting the minimum $MDI_{abs}$ corresponding to a given choice of $b$ and $x$, which occurs at $b = 0.8 \  \mathrm{au}$ and $x = 0.4$. Therefore, the MDI for this choice of step function is zero, representing the closest resemblance to Mercury's CMF. MDI values for different choices of step function are listed in Table~\ref{tab:mdi-cham-cont}. Note that since we perform composition tracking and CMF calculations through post-processing, all step function choices within a single set of simulations (as listed in Table~\ref{tab:simulations_info}) share the same evolved planet masses. Consequently, when comparing MDIs for different choices of $b$ and $x$ within the same group (i.e., \textit{Chambers 2013 continuous}), we are only quantifying how well each case matches Mercury's CMF.

To better reproduce Mercury's CMF, we may prefer closer boundaries with higher $x$ values. However, there is a trade-off: a closer boundary implies less mass available in the inner disk; consequently, higher $x$ values would yield CMF values greater than 1 in this region, which is unphysical. This is why we are not presenting results for higher $x$ values and lower boundaries in Table~\ref{tab:mdi-cham-cont}.

Alternatively, we can select more distant boundaries to allow for higher $x$ values, however, this will result in more total mass in the inner disk, diluting the iron content. Moreover, if the outer disk is depleted of iron beyond a certain level, matching the CMF values of Earth and Venus becomes challenging.

\subsection{Mercury cannot be formed from uniform initial compositions}
\label{sec:combine_results}

\begin{figure}
    \centering
    \includegraphics[width=1\linewidth]{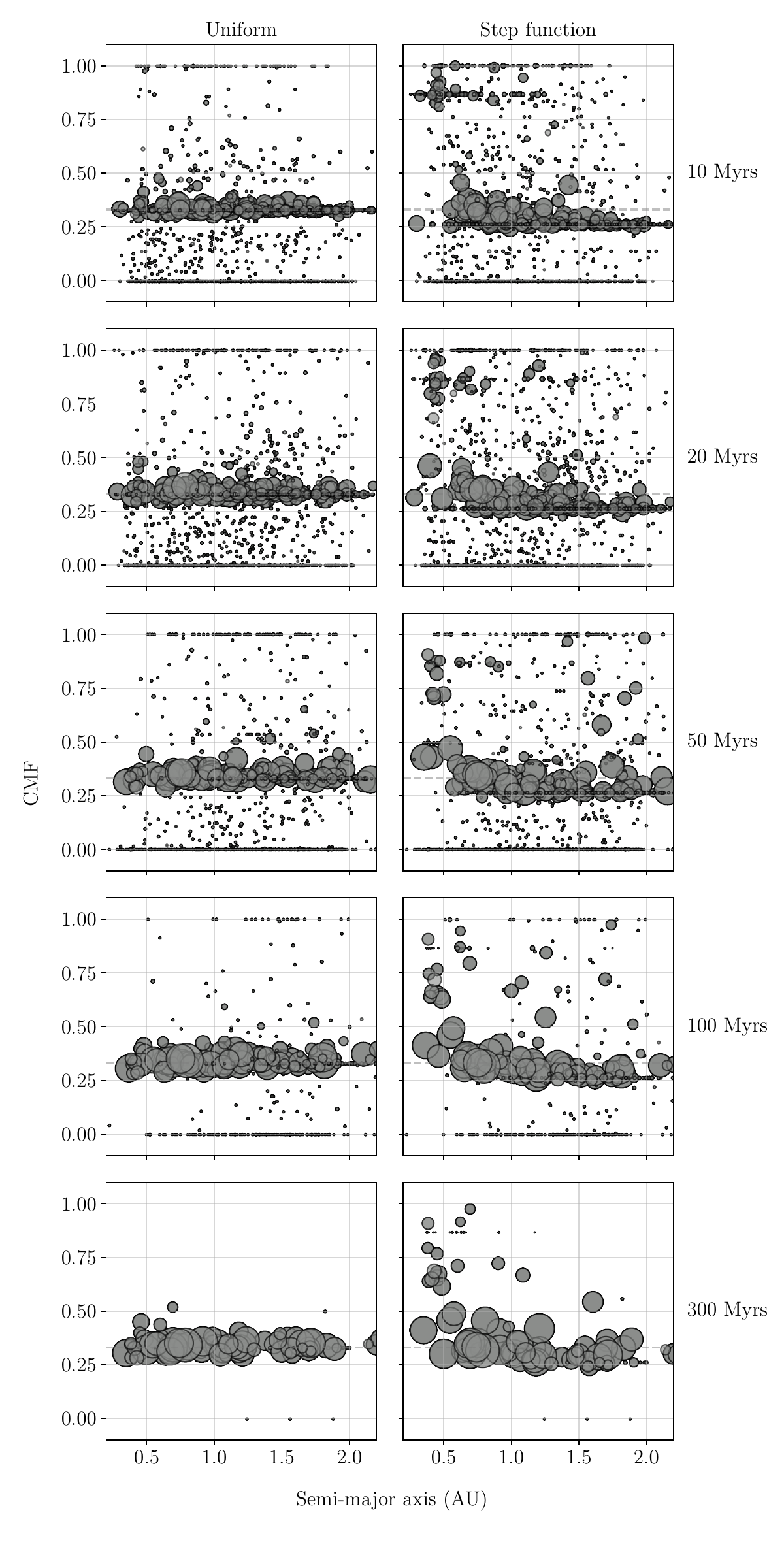}
    \caption{Particles' CMFs versus semi-major axis for different time points during the simulation. Each row corresponds to a different time, starting at 10 Myr at the top and ending at 300 Myr at the bottom. All plots are from \textit{Chambers 2013 continuous} simulation group. The two columns have different initial CMF distributions: the left column starts with a uniform composition ($CMF = 0.33$), and the right column starts with a step function with a 0.7 au boundary and 20\% iron exchange ($b = 0.7, \ x = 0.2$). The evolved planets' CMFs converge toward the planetesimal CMFs from the same region.}
    \label{fig:growing_planets}
\end{figure}

Our results from Sections \ref{sec:uniform_results} and \ref{sec:step_results} indicate that Mercury is unlikely to form from a uniform composition planetesimal disk. This is consistent with previous findings that Mercury formation events are rare \citep{franco_explaining_2022}. We checked our collision statistics to investigate the probability of an energetic giant impact leading to the formation of Mercury. \citet{franco_explaining_2022} report certain ranges for $M_t$, $M_p$, $V_{imp}/V_{esc}$, and impact angle $\theta$ required for Mercury's formation via a single giant impact. Their simulations indicate a probability of less than 1\%. Our results agree with these findings, and we  find an even lower probability, 0.1\% of all our recorded collisions fall within this range. This implies that, apart from the problem of re-accretion and the minimum fragment mass, forming Mercury through a single giant impact is unlikely in our simulations.

However, a Mercury-like planet can evolve from a planetesimal disk whose iron content is parameterized by a step function as a function of orbital distance. This shows that planets' compositions converge to the composition of their parent planetesimals in the same region. Simulations in this work start after planetesimals reach masses around $0.01 M_{\oplus}$. Collisions at earlier stages can lead to effective mantle stripping and spatial separation of iron-rich and iron-poor planetesimals. This corresponds to and justifies our investigated cases of step-function CMFs.

Admittedly, early in the evolution (after a few Myr), erosive collisions produce high-CMF particles. Figure \ref{fig:growing_planets} illustrates the evolution of the two planetesimal disks over time. Each row represents a specific time point, with time increasing from top to bottom. The left column starts with a uniform composition in the disk, and the right column starts with a step function.

High-CMF objects appear at early times in both columns. However, after sufficient time, high-CMF and low-CMF objects merge, and planets converge to the CMF they started with. Therefore, when using a uniform initial composition, planets do not scatter far from their initial state. These results suggest that giant impacts alone, in the late stages of planet formation, are unlikely to form planets similar to Mercury.

Moreover, our simulations do not have any mechanism to remove eroded mantle material. However, small mantle fragments can be depleted through various astrophysical processes \citep{clement_dynamical_2021-1, spalding_solar_2020, melis_rapid_2012, vokrouhlicky_depletion_2000}. In that case, low-CMF objects would not merge with high-CMF objects anymore, and our conclusions may change when considering the depletion mechanism. 

Another important caveat to our results is the minimum fragment mass allowed. Larger mantle fragments are more likely to merge with other objects later on. We discuss this in detail in \S~\ref{sec:min_frag_mass}.

\section{Discussion} \label{sec:discussion}

In this section, we elaborate on the caveats of our simulations. In \S\ref{sec:diss_init_conds}, we address the effects of initial conditions, such as planetesimal surface density distribution. In \S\ref{sec:hyodo}, we examine the effects of initial particle mass on collision statistics. In \S\ref{sec:min_frag_mass}, we discuss our choice of minimum fragment mass and its potential impact on the results. In \S\ref{sec:disk_props}, we discuss the properties of our protoplanetary disk and the differences between solar system and exoplanetary system simulations. Lastly, in \S\ref{sec:big_merc}, we compare the masses of simulated Mercuries with that of the solar system Mercury.

\subsection{Effects of initial conditions} \label{sec:diss_init_conds}

\begin{figure*}
    \centering
    \includegraphics[width=1\linewidth]{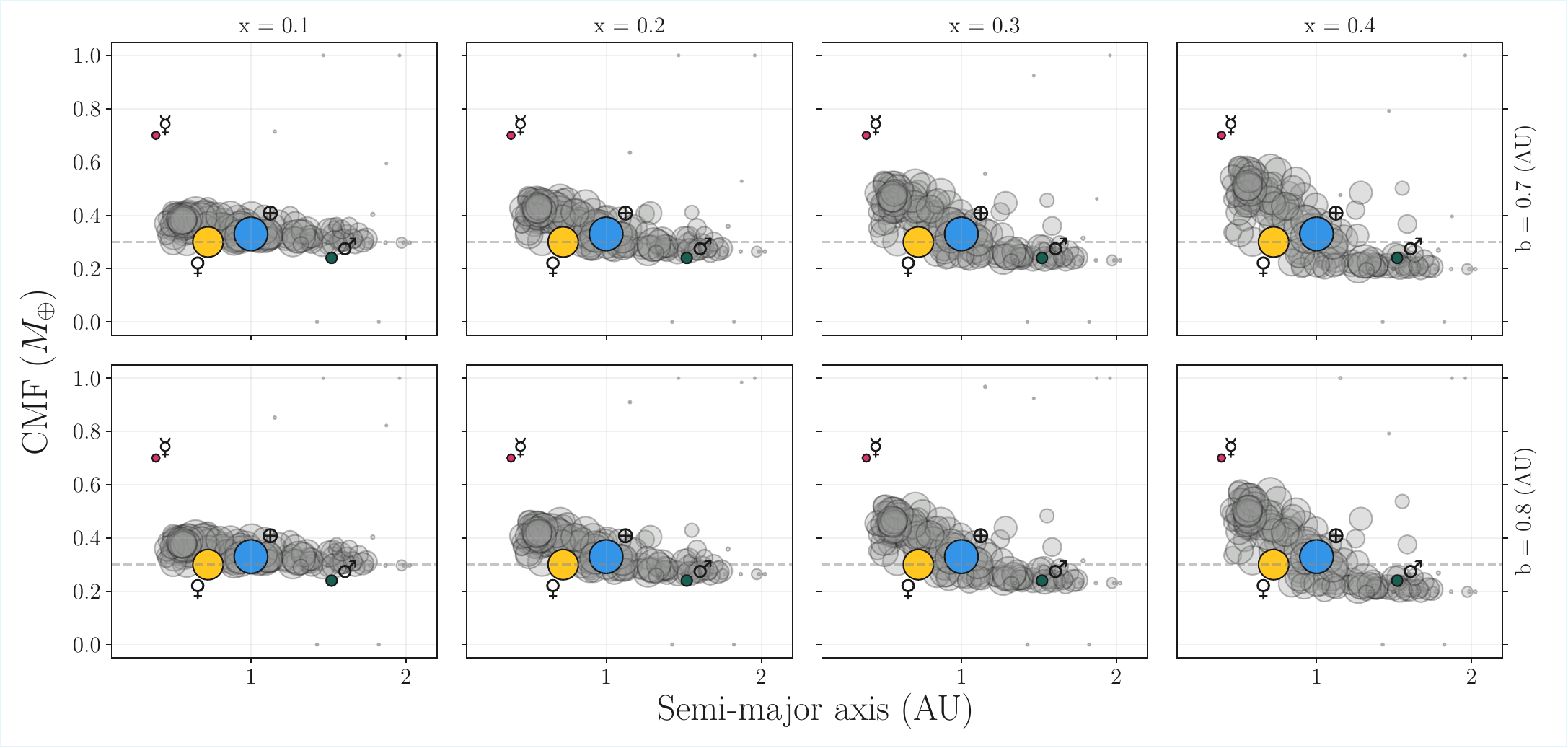}
    \caption{Same as Figure \ref{fig:grid_cham_cont}, but for \textit{Chambers 2013 Bimodal} simulation set.}
    \label{fig:grid-cham-bi-tajer}
\end{figure*}

\begin{figure*}
    \centering
    \includegraphics[width=1\linewidth]{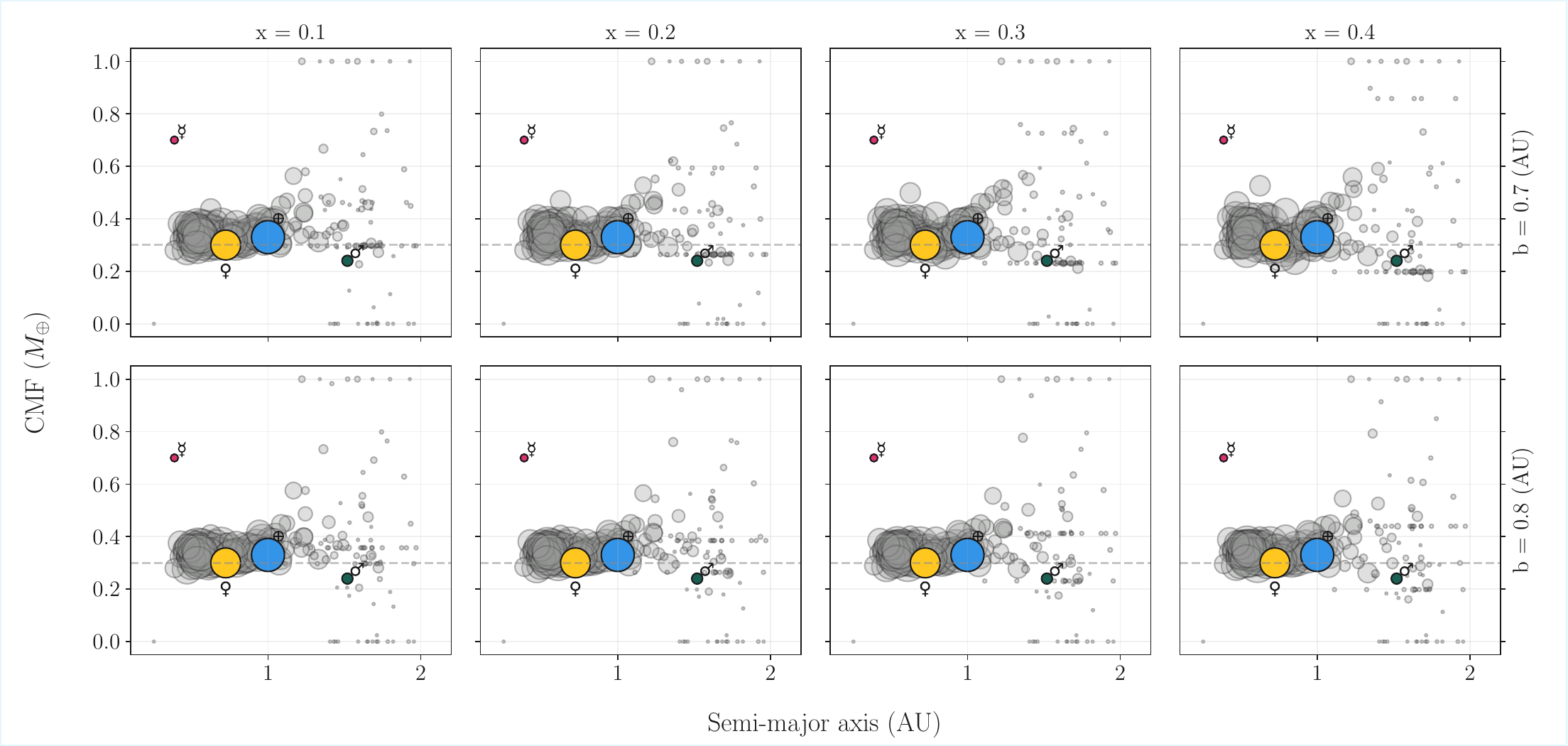}
    \caption{Same as Figure \ref{fig:grid_cham_cont}, but for \textit{Ueda 2021} simulation set.}
    \label{fig:grid-ueda}
\end{figure*}

In this work, we used three different sets of initial surface density distributions, described in detail in \S\ref{subsec: Sigma}, and listed in Table~\ref{tab:simulations_info}. For uniform initial CMF distributions, Figure~\ref{fig:high-metal-all-three} shows the different simulation outcomes for each set. \textit{Ueda 2021} produces many high- and low-CMF objects with relatively low masses, scattered in the outer disk regions ($a > 1.5 \ au$). This may be due to the annulus-like surface density distribution used in this setup. Examining the planetary evolution in \textit{Ueda 2021}, we observe that many newly formed fragments migrate both inward and outward. However, they are more likely to merge with other bodies as they migrate inward, due to the high particle number density. In contrast, fragments that migrate outward encounter lower surface densities and are therefore less likely to interact with other objects, remaining as they were produced during the fragmenting event.

This behavior is not observed in \textit{Chambers 2013 Bimodal}, likely because (1) the annulus effect present in \textit{Ueda 2021} is not applicable here, and (2) the massive planetary embryos in the bimodal distribution exert a stronger gravitational attraction on surrounding bodies, making it more likely for smaller bodies to merge with them. This behavior is also absent in the \textit{Chambers 2013 Continuous} set, likely for the same reason. Although in this case, the planetary embryo masses are proportional to the surface density and there is no bimodal distribution, the embryos are still an order of magnitude more massive, allowing them to attract fragments much more efficiently. 

Figures~\ref{fig:grid-cham-bi-tajer} and~\ref{fig:grid-ueda} show results for \textit{Chambers 2013 bimodal} and \textit{Ueda 2021}, using step-function initial conditions, the same as in Figure~\ref{fig:grid_cham_cont}. In Figure~\ref{fig:grid-cham-bi-tajer}, even bringing more iron to the inner disk cannot fully reproduce Mercury's CMF. With this surface density profile, there is not enough mass available in the inner disk to make the iron swap highly effective. Although the \textit{Chambers 2013 bimodal} simulation set follows the same overall surface density distribution as \textit{Chambers 2013 continuous}, the strong bimodality in this setup (14 planetary embryos with $0.1~M_{\oplus}$ and 140 planetesimals with $0.01~M_{\oplus}$) makes the mass distribution more discrete. Therefore, when choosing smaller boundaries, the inner disk mass is lower than in \textit{Chambers 2013 continuous}, until we reach about 0.7~au, where planetary embryos begin to appear. For example, in the \textit{Chambers 2013 continuous} initial conditions, there are seven objects placed with $a < 0.5~\mathrm{au}$, while in \textit{Chambers 2013 bimodal} there are four objects with $a < 0.5~\mathrm{au}$. However, the total mass in this region for \textit{Chambers 2013 continuous} is $0.12~M_{\oplus}$, whereas for \textit{Chambers 2013 bimodal} it is $0.037~M_{\oplus}$. This is because the continuous and gradual increase in mass in \textit{Chambers 2013 continuous} allows for more objects, and therefore, more total mass.

These differences are small when considering the overall surface density profile but become important when examining Mercury's region. The mass difference between the \textit{Chambers 2013 continuous} and \textit{Chambers 2013 bimodal} sets is large enough to make the iron swap insufficient in the latter case. The sharp changes in mass in this setup imply that we either bring too much iron into the inner disk, where we don't have enough mantle material for swapping, or we move the boundary so far outward that the swapped iron becomes overly diluted. Even if we find a combination of boundary and $x$ that works well for this set, it remains difficult to reproduce Mercury, since there is still not enough mass in Mercury's region. If we move the boundary even further, to have sufficient mass, then the higher-CMF objects will have larger semi-major axes. 

The same scenario is true for the \textit{Ueda 2021} simulation set. Here, the surface density profile is completely different and resembles a log-normal distribution, peaking at $0.78~\mathrm{au}$. Therefore, there is not enough mass in the inner region to make the swap meaningful (Figure~\ref{fig:grid-ueda}).

Moreover, the scattered high- and low-CMF objects in \textit{Ueda 2021} are evident in Figure~\ref{fig:grid-ueda} as well. The results do not change significantly for different sets of boundaries and iron-swap percentages, due to the scarcity of mass in the regions within the chosen boundary.

These results indicate that the choice of initial conditions—specifically, the surface density distribution and the initial masses of planetesimals and planetary embryos—can have important effects on the simulation outcomes. Therefore, to conduct a comprehensive study of Mercury and Exo-Mercury formation, physically informed initial conditions must be carefully considered.

\subsection{Erosive collisions are more efficient with smaller bodies} \label{sec:hyodo}

\begin{figure}
    \centering
    \includegraphics[width=1\linewidth]{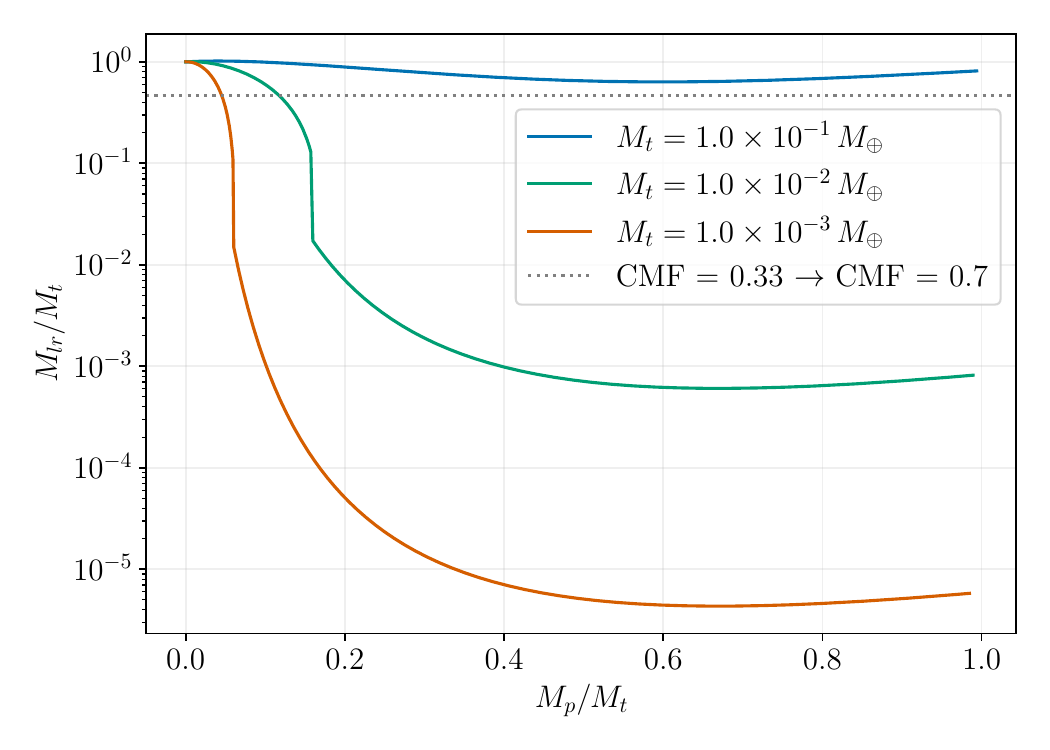}
   \caption{Mass of the largest remnant to target mass ratio versus projectile to target mass ratio. Larger ratios promote more mantle stripping; however, it is still difficult to erode mantle from a high-mass target.}
    \label{fig:mlr_mt_gamma}
\end{figure}

\citet{hyodo_modification_2020} show that collisions between small planetesimals ($10^{18} - 10^{25}$ g $\approx 10^{-10} - 10^{-3} M_{\oplus}$) are efficient enough to erode mantle material, leaving iron-enriched bodies that can later merge into Mercury-like planets. To have an erosive event, the impact velocity $V_{imp}$ should exceed the escape velocity $V_{esc}$. The higher the $V_{imp}/V_{esc}$ ratio, the more mass can be eroded during the impact.

In Mercury's region at $\approx 0.35$ au, Keplerian velocities ($V_k$) are higher than in more distant regions, such as Earth's. \citet{hyodo_modification_2020} derive the impact velocity from the random velocity $V_{ran}$:

\begin{equation}
    V_{ran} = \sqrt{e^2 + i^2}v_k 
\end{equation}
\begin{equation}
    V_{imp} = \sqrt{V_{esc}^2 + V_{ran}^2}
\end{equation}

Where $e$ and $i$ are the eccentricity and inclination of the orbit, both assumed to be low. Moreover, in Mercury's region, particles' orbits can be perturbed by distant massive objects, leading to higher eccentricities. These effects increase the efficiency of mantle stripping in Mercury's region \citep{hyodo_modification_2020}.

The transition between a merging and an erosive event happens when the impact velocity is higher than the escape velocity \citep{Leinhardt2012}. Since the escape velocity is proportional to the square root of mass ($V_{esc} = \sqrt{2GM/R}$), collisions between lower mass objects are more likely to exceed the escape velocity.

In the collision prescription derived from \citet{Leinhardt2012} (See \S\ref{subsec: N-body simulations}), the mass of the largest remnant from a collision ($M_{lr}$) is a function of the impact energy, which itself depends on the impact velocity, target and projectile masses ($M_{t}$ and $M_{p}$), and the impact angle.

We plot changes in $M_{lr}/M_{t}$ with varying projectile to target mass ratio $M_p/M_t$. Here, we assume a head-on impact, with a Keplerian velocity in Mercury's region ($a \approx 0.35$ au), an eccentricity of 0.2, and an inclination of 0. The dotted line indicates the $M_{lr}/M_{t}$ required for an object with $CMF = 0.33$ to reach a CMF of 0.7. Any $M_{lr}/M_{t}$ below this threshold corresponds to more mantle erosion, producing a largest remnant with a higher CMF. Curves for lower mass targets fall well below the dotted line, for a large range of projectile to target mass ratios. However, for a larger target ($M_t = 1.0 \times 10^{-1} M_{\oplus}$), $M_{lr}/M_{t}$ never falls below the dotted line, meaning that a target with this mass is unable to reach a 0.7 CMF within a single impact.

In our simulations, we start with a disk of planetesimals that are too large to undergo major erosive collisions. Due to computational costs, we cannot make the planetesimals smaller. However, we consider the results of \citet{hyodo_modification_2020} as one possible pathway to form an iron-rich planetesimal region closer to the host star. We assume that highly erosive collisions have occurred before the starting point of our simulations. As discussed above, erosive collisions act more efficiently in Mercury's region. Consequently, the inner region contains more iron-enriched bodies, motivating our choice of a step-function initial CMF distribution.

\subsection{Minimum fragment mass}
\label{sec:min_frag_mass}

In our simulations, we need to define a minimum fragment mass $M_{min, frag}$ for two reasons. First, because we need to have a reference point for fragment masses. Equations derived from \citet{Leinhardt2012} compute the mass of the largest remnant, but the remaining mass can be distributed into fragments of arbitrary size. Secondly, setting a very small $M_{min,frag}$ will lead to a high number of fragments, adding up to the number of bodies in the simulation. This will slow the simulation down extensively. 

Accordingly, we arbitrarily define $M_{min,frag}$ as half the mass of planetesimals in the \textit{Chambers 2013 bimodal} simulation set ($M_{min,frag} = 1/2 \times 0.01 M_{\oplus}$). The values of $M_{min,frag}$ for the other simulation sets are listed in Table \ref{tab:simulations_info}.


Because of this limitation in the minimum fragment mass allowed, if an erosive collision is not energetic enough to erode a mass larger than $M_{min,frag}$, we need to resolve the collision in another way. Depending on the collision parameters, such collisions might be treated as mergers \citep{childs_collisional_2022}. Therefore, some erosive collisions are inevitably ignored.

Moreover, higher mass also implies stronger gravitational effects, therefore higher mass fragments are more likely to merge with other objects. As shown in Fig. \ref{fig:growing_planets}, high- and low-CMF objects appear early in the evolution but eventually merge with one another.

Using a lower minimum fragment mass will 1) possibly result in more fragmentation, since the code will be able to resolve more erosive collisions, 2) will reduce the likelihood of high- and low-CMF fragments merging, and might alter the convergence to the initial CMFs.


\begin{figure*}
    \centering
    \includegraphics[width=1\linewidth]{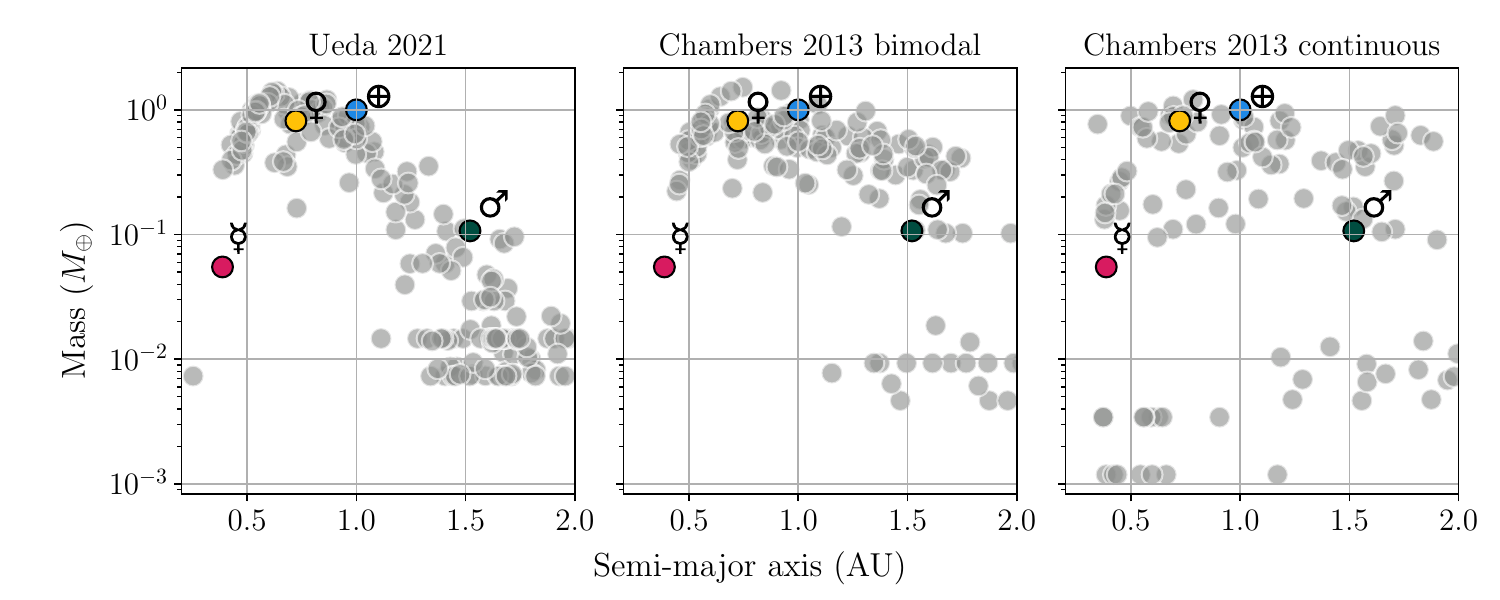}
    \caption{Mass versus semi-major axis for final planets derived from each simulation set. Solar system planets are plotted and annotated with their Roman letters.}
    \label{fig:mass-semi}
\end{figure*}

\subsection{Do the solar system and exoplanetary systems share similar planetesimal disk properties?}
\label{sec:disk_props}

Reproducing all terrestrial planets in the solar system is not within the scope of this project. However, we use a protoplanetary disk based on solar system models, even though our aim is to study the probability of forming Mercury-like planets both in the solar system and among stars.

To do this, we also need to model exoplanetary systems. \cite{chiang_minimum-mass_2013} show that, based on the Kepler planets, exoplanetary systems could have protoplanetary disks up to five times more massive than the minimum mass solar nebula. Accordingly, to study exo-Mercuries, it is necessary to conduct simulations with more massive disks. However, higher disk mass implies a higher number of bodies, increasing computational cost to a level that makes it unfeasible with the tools developed in this work. Including fragmentation adds even more bodies to the simulation, making it even more time-consuming. It is possible to increase the mass of each body instead of their number; however, this ignores the important earlier phases, where erosive collisions are more efficient.

Another challenge in modeling exoplanetary systems is the edges of the disk. In the Kepler systems studied by \cite{chiang_minimum-mass_2013}, numerous short-period planets are present, with orbital periods on the order of a few days. If these planets formed in situ, and if we want to set the inner edges of the initial disk so close to the host star, we need to choose a timestep small enough to capture changes in the shortest orbits. Reducing the timestep also slows down the simulations, so we either need to use faster integrators \citep[e.g., TRACE][]{lu_trace_2024} or GPU-based N-body codes. Both of these approaches will be explored in our upcoming work.

\subsection{Simulated Mercuries are too big}
\label{sec:big_merc}

We see a clear trend of forming both Mercury-like and Earth-like planets in most of the simulations (Fig. \ref{fig:mass-semi}). However, we cannot match Mercury's mass—the final simulated inner planets are significantly more massive than the solar system Mercury. This is a persistent problem in many solar system simulations attempting to reproduce all terrestrial planets \citep{clement_dynamical_2021}. However, reproducing terrestrial planets accurately is not within the scope of this project. In this work, we focus only on the probability of forming a high-CMF planet close to the star, along with lower-CMF planets in the same system.

It is expected that using the planetesimal surface density distribution from \citet{ueda_early_2021} helps with creating lower mass Mercuries. The log-normal distribution implies there is a smaller mass reservoir both in Mercury's and Mars's regions. In the \textit{Ueda 2021} simulation set, we see lower mass objects with high CMFs, in Mars's proximity. However, this is not the case for the inner regions. This might be due to the dynamical properties of Mercury's region. Collisions are more likely to happen in a closer proximity to the star, therefore, it is more likely for fragments to merge and create more massive bodies.

\section{Conclusions}
\label{sec:conclusions}

In this study, we investigated whether Mercury's high CMF can be achieved from giant erosive impacts in a uniform composition disk during the planetesimal stage, or if some process is needed to enrich the inner disk with iron. We conducted N-body simulations of planet formation, starting at a stage where planetesimals are differentiated. We included fragmentation and tracked the CMF of bodies through their evolution. We used three different simulation sets with various initial surface density distributions. Our main conclusions are:

\begin{itemize}
    \item It is unlikely to form a Mercury-like planet from a disk of planetesimals with a uniform Earth-like composition (i.e., CMF = 0.33). Figure \ref{fig:high-metal-all-three} illustrates simulation results using uniform initial CMF distributions. Even though fragmentation will produce high-CMF objects through the planets' evolution, in the absence of a process to deplete eroded mantle material, high- and low-CMF objects will eventually merge, converging to their initial composition (see Figure~\ref{fig:growing_planets}). 
\item It is possible to achieve a Mercury-level CMF of ~70\% from a planetesimal disk with an iron-rich (i.e., high-CMF) inner region, and a lower-CMF outer region. Figure~\ref{fig:grid_cham_cont} depicts simulation results when using step function initial CMF distributions. This setup is motivated by previous studies, suggesting that iron-rich planetesimals can form in the proximity of the Sun \citep{johansen_nucleation_2022, hyodo_modification_2020, kruss_seeding_2018, Kruss2018, Wurm2018, wurm_photophoretic_2013}.
    \item To accurately simulate effects of fragmentation in the formation of iron-rich planets, higher resolution simulations with smaller mass objects are needed. This is important because 1) erosive collisions are more efficient between smaller mass particles, and 2) allowing for lower values of minimum fragment mass will enable capturing more erosive collisions. It is also important to account for astrophysical processes that can deplete the eroded mantle material. All of these limitations can be addressed in future work.

\end{itemize}

\section{Acknowledgments}

Haniyeh Tajer thanks the LSST-DA Data Science Fellowship Program, which is funded by LSST-DA, the Brinson Foundation, the WoodNext Foundation, and the Research Corporation for Science Advancement Foundation; her participation in the program has benefited this work.

\clearpage

\bibliography{sample631}{}

\begin{thebibliography}{}
\expandafter\ifx\csname natexlab\endcsname\relax\def\natexlab#1{#1}\fi
\providecommand{\url}[1]{\href{#1}{#1}}
\providecommand{\dodoi}[1]{doi:~\href{http://doi.org/#1}{\nolinkurl{#1}}}
\providecommand{\doeprint}[1]{\href{http://ascl.net/#1}{\nolinkurl{http://ascl.net/#1}}}
\providecommand{\doarXiv}[1]{\href{https://arxiv.org/abs/#1}{\nolinkurl{https://arxiv.org/abs/#1}}}

\bibitem[{Adibekyan {et~al.}(2021)Adibekyan, Dorn, Sousa, Santos, Bitsch, Israelian, Mordasini, Barros, Delgado~Mena, Demangeon, Faria, Figueira, Hakobyan, Oshagh, Soares, Kunitomo, Takeda, Jofré, Petrucci, \& Martioli}]{adibekyan_compositional_2021}
Adibekyan, V., Dorn, C., Sousa, S.~G., {et~al.} 2021, Science, 374, 330, \dodoi{10.1126/science.abg8794}

\bibitem[{{Adibekyan} {et~al.}(2022){Adibekyan}, {Dorn}, {Sousa}, {Santos}, {Bitsch}, {Israelian}, {Mordasini}, {Barros}, {Delgado Mena}, {Demangeon}, {Faria}, {Figueira}, {Hakobyan}, {Osagh}, {Soares}, {Kunitomo}, {Takeda}, {Jofr{\'e}}, {Petrucc}, \& {Martioli}}]{adibekyan2022}
{Adibekyan}, V., {Dorn}, C., {Sousa}, S., {et~al.} 2022, in European Planetary Science Congress, EPSC2022--74, \dodoi{10.5194/epsc2022-74}

\bibitem[{Benz {et~al.}(2007)Benz, Anic, Horner, \& Whitby}]{benz_origin_2007}
Benz, W., Anic, A., Horner, J., \& Whitby, J.~A. 2007, Space Science Reviews, 132, 189, \dodoi{10.1007/s11214-007-9284-1}

\bibitem[{Benz {et~al.}(1988)Benz, Slattery, \& Cameron}]{benz_collisional_1988}
Benz, W., Slattery, W.~L., \& Cameron, A. G.~W. 1988, Icarus, 74, 516, \dodoi{10.1016/0019-1035(88)90118-2}

\bibitem[{Bonomo {et~al.}(2019)Bonomo, Zeng, Damasso, Leinhardt, Justesen, Lopez, Lund, Malavolta, Aguirre, Buchhave, Corsaro, Denman, Lopez-Morales, Mills, Mortier, Rice, Sozzetti, Vanderburg, Affer, Arentoft, Benbakoura, Bouchy, Christensen-Dalsgaard, Cameron, Cosentino, Dressing, Dumusque, Figueira, Fiorenzano, García, Handberg, Harutyunyan, Johnson, Kjeldsen, Latham, Lovis, Lundkvist, Mathur, Mayor, Micela, Molinari, Motalebi, Nascimbeni, Nava, Pepe, Phillips, Piotto, Poretti, Sasselov, Ségransan, Udry, \& Watson}]{bonomo_giant_2019}
Bonomo, A.~S., Zeng, L., Damasso, M., {et~al.} 2019, Nature Astronomy, 3, 416, \dodoi{10.1038/s41550-018-0684-9}

\bibitem[{Chambers(2013)}]{Chambers2013}
Chambers, J. 2013, Icarus, 224, 43, \dodoi{10.1016/j.icarus.2013.02.015}

\bibitem[{Chau {et~al.}(2018)Chau, Reinhardt, Helled, \& Stadel}]{chau_forming_2018}
Chau, A., Reinhardt, C., Helled, R., \& Stadel, J. 2018, The Astrophysical Journal, 865, 35, \dodoi{10.3847/1538-4357/aad8b0}

\bibitem[{Chiang \& Laughlin(2013)}]{chiang_minimum-mass_2013}
Chiang, E., \& Laughlin, G. 2013, Monthly Notices of the Royal Astronomical Society, 431, 3444, \dodoi{10.1093/mnras/stt424}

\bibitem[{Childs \& Steffen(2022)}]{childs_collisional_2022}
Childs, A.~C., \& Steffen, J.~H. 2022, Monthly Notices of the Royal Astronomical Society, 511, 1848, \dodoi{10.1093/mnras/stac158}

\bibitem[{Christiansen {et~al.}(2025)Christiansen, McElroy, Harbut, Ciardi, Crane, Good, Hardegree-Ullman, Kesseli, Lund, Lynn, Muthiar, Nilsson, Oluyide, Papin, Rivera, Swain, Susemiehl, Tam, Eyken, \& Beichman}]{christiansen_nasa_2025}
Christiansen, J.~L., McElroy, D.~L., Harbut, M., {et~al.} 2025, The {NASA} {Exoplanet} {Archive} and {Exoplanet} {Follow}-up {Observing} {Program}: {Data}, {Tools}, and {Usage},  arXiv, \dodoi{10.48550/arXiv.2506.03299}

\bibitem[{Clement \& Chambers(2021)}]{clement_dynamical_2021-1}
Clement, M.~S., \& Chambers, J.~E. 2021, The Astronomical Journal, 162, 3, \dodoi{10.3847/1538-3881/abfb6c}

\bibitem[{Clement {et~al.}(2021)Clement, Chambers, \& Jackson}]{clement_dynamical_2021}
Clement, M.~S., Chambers, J.~E., \& Jackson, A.~P. 2021, The Astronomical Journal, 161, 240, \dodoi{10.3847/1538-3881/abf09f}

\bibitem[{Clement {et~al.}(2019)Clement, Kaib, \& Chambers}]{Clement2019}
Clement, M.~S., Kaib, N.~A., \& Chambers, J.~E. 2019, The Astronomical Journal, 157, 208, \dodoi{10.3847/1538-3881/ab164f}

\bibitem[{Dou {et~al.}(2024)Dou, Carter, \& Leinhardt}]{10.1093/mnras/stae644}
Dou, J., Carter, P.~J., \& Leinhardt, Z.~M. 2024, Monthly Notices of the Royal Astronomical Society, 529, 2577, \dodoi{10.1093/mnras/stae644}

\bibitem[{Ferich {et~al.}(2024)Ferich, Childs, \& Steffen}]{ferich2024}
Ferich, N., Childs, A.~C., \& Steffen, J.~H. 2024, arXiv

\bibitem[{Franco {et~al.}(2022)Franco, Izidoro, Winter, Torres, \& Amarante}]{franco_explaining_2022}
Franco, P., Izidoro, A., Winter, O.~C., Torres, K.~S., \& Amarante, A. 2022, Monthly Notices of the Royal Astronomical Society, 515, 5576, \dodoi{10.1093/mnras/stac2183}

\bibitem[{Hinkel {et~al.}(2014)Hinkel, Timmes, Young, Pagano, \& Turnbull}]{hinkel_stellar_2014}
Hinkel, N.~R., Timmes, F.~X., Young, P.~A., Pagano, M.~D., \& Turnbull, M.~C. 2014, The Astronomical Journal, 148, 54, \dodoi{10.1088/0004-6256/148/3/54}

\bibitem[{Hyodo {et~al.}(2020)Hyodo, Genda, \& Brasser}]{hyodo_modification_2020}
Hyodo, R., Genda, H., \& Brasser, R. 2020, arXiv, \dodoi{10.48550/arxiv.2008.08490}

\bibitem[{Johansen \& Dorn(2022)}]{johansen_nucleation_2022}
Johansen, A., \& Dorn, C. 2022, arXiv, \dodoi{10.48550/arxiv.2204.04241}

\bibitem[{Kruss \& Wurm(2018)}]{kruss_seeding_2018}
Kruss, M., \& Wurm, G. 2018, arXiv, \dodoi{10.48550/arxiv.1812.05338}

\bibitem[{{Kruss} \& {Wurm}(2018)}]{Kruss2018}
{Kruss}, M., \& {Wurm}, G. 2018, \apj, 869, 45, \dodoi{10.3847/1538-4357/aaec78}

\bibitem[{Kruss \& Wurm(2020)}]{kruss_composition_2020}
Kruss, M., \& Wurm, G. 2020, arXiv, \dodoi{10.48550/arxiv.2006.13692}

\bibitem[{Leinhardt \& Stewart(2012)}]{Leinhardt2012}
Leinhardt, Z.~M., \& Stewart, S.~T. 2012, The Astrophysical Journal, 745, 79, \dodoi{10.1088/0004-637x/745/1/79}

\bibitem[{Lu {et~al.}(2024)Lu, Hernandez, \& Rein}]{lu_trace_2024}
Lu, T., Hernandez, D.~M., \& Rein, H. 2024, Monthly Notices of the Royal Astronomical Society, 533, 3708, \dodoi{10.1093/mnras/stae1982}

\bibitem[{{Marcus} {et~al.}(2010){Marcus}, {Sasselov}, {Hernquist}, \& {Stewart}}]{Marcus2010}
{Marcus}, R.~A., {Sasselov}, D., {Hernquist}, L., \& {Stewart}, S.~T. 2010, \apjl, 712, L73, \dodoi{10.1088/2041-8205/712/1/L73}

\bibitem[{Marcus {et~al.}(2010)Marcus, Sasselov, Stewart, \& Hernquist}]{marcus_watericy_2010}
Marcus, R.~A., Sasselov, D., Stewart, S.~T., \& Hernquist, L. 2010, The Astrophysical Journal Letters, 719, L45, \dodoi{10.1088/2041-8205/719/1/l45}

\bibitem[{Melis {et~al.}(2012)Melis, Zuckerman, Rhee, Song, Murphy, \& Bessell}]{melis_rapid_2012}
Melis, C., Zuckerman, B., Rhee, J.~H., {et~al.} 2012, Nature, 487, 74, \dodoi{10.1038/nature11210}

\bibitem[{{Peplowski} {et~al.}(2011){Peplowski}, {Evans}, {Hauck}, {McCoy}, {Boynton}, {Gillis-Davis}, {Ebel}, {Goldsten}, {Hamara}, {Lawrence}, {McNutt}, {Nittler}, {Solomon}, {Rhodes}, {Sprague}, {Starr}, \& {Stockstill-Cahill}}]{Peplowski2011}
{Peplowski}, P.~N., {Evans}, L.~G., {Hauck}, S.~A., {et~al.} 2011, Science, 333, 1850, \dodoi{10.1126/science.1211576}

\bibitem[{Rein \& Liu(2012)}]{rebound2012}
Rein, H., \& Liu, S.-F. 2012, Astronomy \& Astrophysics, 537, A128, \dodoi{10.1051/0004-6361/201118085}

\bibitem[{Rein {et~al.}(2019)Rein, Hernandez, Tamayo, Brown, Eckels, Holmes, Lau, Leblanc, \& Silburt}]{rein_hybrid_2019}
Rein, H., Hernandez, D.~M., Tamayo, D., {et~al.} 2019, Monthly Notices of the Royal Astronomical Society, 485, 5490, \dodoi{10.1093/mnras/stz769}

\bibitem[{Santerne {et~al.}(2018)Santerne, Brugger, Armstrong, Adibekyan, Lillo-Box, Gosselin, Aguichine, Almenara, Barrado, Barros, Bayliss, Boisse, Bonomo, Bouchy, Brown, Deleuil, Mena, Demangeon, Díaz, Doyle, Dumusque, Faedi, Faria, Figueira, Foxell, Giles, Hébrard, Hojjatpanah, Hobson, Jackman, King, Kirk, Lam, Ligi, Lovis, Louden, McCormac, Mousis, Neal, Osborn, Pepe, Pollacco, Santos, Sousa, Udry, \& Vigan}]{santerne_earth-sized_2018}
Santerne, A., Brugger, B., Armstrong, D.~J., {et~al.} 2018, Nature Astronomy, 2, 393, \dodoi{10.1038/s41550-018-0420-5}

\bibitem[{Schulze {et~al.}(2021)Schulze, Wang, Johnson, Gaudi, Unterborn, \& Panero}]{schulze_probability_2021}
Schulze, J.~G., Wang, J., Johnson, J.~A., {et~al.} 2021, The Planetary Science Journal, 2, 113, \dodoi{10.3847/psj/abcaa8}

\bibitem[{Sohl \& Schubert(2015)}]{SOHL201523}
Sohl, F., \& Schubert, G. 2015, in Treatise on Geophysics (Second Edition), second edition edn., ed. G.~Schubert (Oxford: Elsevier), 23--64, \dodoi{https://doi.org/10.1016/B978-0-444-53802-4.00166-4}

\bibitem[{Spalding \& Adams(2020)}]{spalding_solar_2020}
Spalding, C., \& Adams, F.~C. 2020, The Planetary Science Journal, 1, 7, \dodoi{10.3847/PSJ/ab781f}

\bibitem[{Taylor \& McLennan(2008)}]{Taylor_McLennan_2008}
Taylor, S.~R., \& McLennan, S. 2008, The planets: their formation and differentiation, Cambridge Planetary Science (Cambridge University Press), 5–31

\bibitem[{Toledo-Padrón {et~al.}(2020)Toledo-Padrón, Lovis, Mascareño, Barros, Hernández, Sozzetti, Bouchy, Osorio, Rebolo, Cristiani, Pepe, Santos, Sousa, Tabernero, Lillo-Box, Bossini, Adibekyan, Allart, Damasso, D’Odorico, Figueira, Lavie, Curto, Mehner, Micela, Modigliani, Nunes, Pallé, Abreu, Affolter, Alibert, Aliverti, Prieto, Alves, Amate, Avila, Baldini, Bandy, Benatti, Benz, Bianco, Broeg, Cabral, Calderone, Cirami, Coelho, Conconi, Coretti, Cumani, Cupani, Deiries, Dekker, Delabre, Demangeon, Marcantonio, Ehrenreich, Fragoso, Genolet, Genoni, Santos, Hughes, Iwert, Knudstrup, Landoni, Lizon, Maire, Manescau, Martins, Mégevand, Molaro, Monteiro, Monteiro, Moschetti, Mueller, Oggioni, Oliveira, Oshagh, Pariani, Pasquini, Poretti, Rasilla, Redaelli, Riva, Tschudi, Santin, Santos, Segovia, Sosnowska, Spanò, Tenegi, Udry, Zanutta, \& Zerbi}]{toledo-padron_characterization_2020}
Toledo-Padrón, B., Lovis, C., Mascareño, A.~S., {et~al.} 2020, Astronomy \& Astrophysics, 641, A92, \dodoi{10.1051/0004-6361/202038187}

\bibitem[{Ueda {et~al.}(2021)Ueda, Ogihara, Kokubo, \& Okuzumi}]{ueda_early_2021}
Ueda, T., Ogihara, M., Kokubo, E., \& Okuzumi, S. 2021, The Astrophysical Journal, 921, L5, \dodoi{10.3847/2041-8213/ac2f3b}

\bibitem[{Unterborn {et~al.}(2023)Unterborn, Desch, Haldemann, Lorenzo, Schulze, Hinkel, \& Panero}]{Unterborn_2023}
Unterborn, C.~T., Desch, S.~J., Haldemann, J., {et~al.} 2023, The Astrophysical Journal, 944, 42, \dodoi{10.3847/1538-4357/acaa3b}

\bibitem[{Valencia {et~al.}(2006)Valencia, O'Connell, \& Sasselov}]{valencia_internal_2006}
Valencia, D., O'Connell, R.~J., \& Sasselov, D. 2006, Icarus, 181, 545, \dodoi{10.1016/j.icarus.2005.11.021}

\bibitem[{Vokrouhlický {et~al.}(2000)Vokrouhlický, Farinella, \& Bottke}]{vokrouhlicky_depletion_2000}
Vokrouhlický, D., Farinella, P., \& Bottke, W.~F. 2000, Icarus, 148, 147, \dodoi{10.1006/icar.2000.6468}

\bibitem[{Walsh {et~al.}(2011)Walsh, Morbidelli, Raymond, O'Brien, \& Mandell}]{Walsh2011}
Walsh, K.~J., Morbidelli, A., Raymond, S.~N., O'Brien, D.~P., \& Mandell, A.~M. 2011, Nature, 475, 206, \dodoi{10.1038/nature10201}

\bibitem[{{Wurm}(2018)}]{Wurm2018}
{Wurm}, G. 2018, Geosciences, 8, 310, \dodoi{10.3390/geosciences8090310}

\bibitem[{Wurm {et~al.}(2013)Wurm, Trieloff, \& Rauer}]{wurm_photophoretic_2013}
Wurm, G., Trieloff, M., \& Rauer, H. 2013, arXiv, \dodoi{10.48550/arxiv.1305.0689}

\bibitem[{Zsom {et~al.}(2010)Zsom, Ormel, Güttler, Blum, \& Dullemond}]{zsom_outcome_2010}
Zsom, A., Ormel, C.~W., Güttler, C., Blum, J., \& Dullemond, C.~P. 2010, Astronomy \& Astrophysics, 513, A57, \dodoi{10.1051/0004-6361/200912976}

\end{thebibliography}
\bibliographystyle{aasjournal}



\begin{acknowledgments}
\end{acknowledgments}

%






\appendix

\section{Initial CMFs for different step functions}
\label{sec:app_tables}

\begin{table*}[h!]
\centering
\caption{Initial CMFs for \textit{Chambers 2013 bimodal} simulation set.}
\label{tab:cmf_grid_data_cham_bench}
\begin{tabular}{c  c  c  c}
\toprule
CMF boundary ($b$) [au] & Percentage of iron exchange ($x$) & Inner CMFs & Outer CMFs \\
\midrule
0.6    & 0.05        & 0.73       & 0.31       \\
0.7    & 0.1         & 0.43       & 0.30       \\
0.7    & 0.2         & 0.52       & 0.26       \\
0.7    & 0.3         & 0.62       & 0.23       \\
0.7    & 0.4         & 0.72       & 0.20       \\
0.7    & 0.5         & 0.82       & 0.17       \\
0.8    & 0.1         & 0.41       & 0.30       \\
0.8    & 0.2         & 0.49       & 0.26       \\
0.8    & 0.3         & 0.57       & 0.23       \\
0.8    & 0.4         & 0.65       & 0.20       \\
0.8    & 0.5         & 0.74       & 0.17       \\

\bottomrule
\end{tabular}
\end{table*}

\begin{table*}[h!]
\centering
\caption{Initial CMFs for \textit{Chambers 2013 continuous} simulation set.}
\label{tab:cmf_grid_data_cham_bench}
\begin{tabular}{c  c  c  c}
\toprule
CMF boundary ($b$) [au] & Percentage of iron exchange ($x$) & Inner CMFs & Outer CMFs \\
\midrule
0.5    & 0.1         & 0.99       & 0.30       \\
0.6    & 0.1         & 0.77       & 0.30       \\
0.7    & 0.1         & 0.60       & 0.30       \\
0.7    & 0.2         & 0.87       & 0.26       \\
0.8    & 0.1         & 0.44       & 0.30       \\
0.8    & 0.2         & 0.54       & 0.26       \\
0.8    & 0.3         & 0.65       & 0.23       \\
0.8    & 0.4         & 0.76       & 0.20       \\
\bottomrule
\end{tabular}
\end{table*}

\begin{table*}[h!]
\centering
\caption{Initial CMFs for \textit{Ueda 2021} simulation set.}
\label{tab:cmf_grid_data_ueda}
\begin{tabular}{c  c  c  c}
\toprule
CMF boundary ($b$) [au] & Percentage of iron exchange ($x$) & Inner CMFs & Outer CMFs \\
\midrule

0.7    & 0.1         & 0.46       & 0.30       \\
0.7    & 0.2         & 0.59       & 0.26       \\
0.7    & 0.3         & 0.73       & 0.23       \\
0.7    & 0.4         & 0.86       & 0.20       \\
0.8    & 0.1         & 0.36       & 0.30       \\
0.8    & 0.2         & 0.38       & 0.26       \\
0.8    & 0.3         & 0.41       & 0.23       \\
0.8    & 0.4         & 0.44       & 0.20       \\

\bottomrule
\end{tabular}
\end{table*}

\clearpage

\end{CJK*}
\end{document}